\documentclass[12pt]{article}
\usepackage{amsmath,amsthm,amssymb,epic,eepic,texdraw}
\newcommand{\nc}{\newcommand}
\nc{\mref}[1]{(\ref{#1})}
\nc{\vt}{v_{2\gL_0}}
\nc{\vo}{v_{\gL_0}}
\nc{\vot}{v_{\gL_1+\gL_0}}
\nc{\vw}{v_{\gL_1}}
\nc{\ppmm}{\genfrac{}{}{-10pt}{10pt}{++}{--}}
\nc{\wom}[5]{\Omega\left(\left.\begin{array}{ll}{#1}&{#2}\\{#3}&{#4}\end{array}\right|{#5}\right)}
\nc{\com}[5]{\chi\left(\left.\begin{array}{ll}{#1}&{#2}\\{#3}&{#4}\end{array}\right|{#5}\right)}
\nc{\we}[5]{W\left(\left.\begin{array}{ll}{#1}&{#2}\\{#3}&{#4}\end{array}\right|{#5}\right)}
\nc{\cet}[7]{C^{#6}_{#7}\left(\left.\begin{array}{ll}{#1}&{#2}\\{#3}&{#4}\end{array}\right|{#5}\right)}
\nc{\bcet}[7]{\bar{C}^{#6}_{#7}\left(\left.\begin{array}{ll}{#1}&{#2}\\{#3}&{#4}\end{array}\right|{#5}\right)}
\nc{\wet}[7]{W^{#6}_{#7}\left(\left.\begin{array}{ll}{#1}&{#2}\\{#3}&{#4}\end{array}\right|{#5}\right)}
\nc{\bwet}[7]{\overline{W}^{#6}_{#7}\left(\left.\begin{array}{ll}{#1}&{#2}\\{#3}&{#4}\end{array}\right|{#5}\right)}
\nc{\wec}[7]{\widetilde{W}^{#6}_{#7}\left(\left.\begin{array}{ll}{#1}&{#2}\\{#3}&{#4}\end{array}\right|{#5}\right)}
\nc{\qbinom}[2]{{\genfrac{[}{]}{0pt}{}{{#1}}{{#2}}}_{q}}
\nc{\hg}[4]{{}_2\phi_1\left({{{#1}\,\,\,{#2}}\atop{{#3}}};p,
                     {#4}\right)}
\nc{\hhg}[4]{\phi\left({{{#1}\,\,\,{#2}}\atop{{#3}}};
                     {#4}\right)}
\nc{\fullhhg}[5]{{_2}\phi_1\left({{{#1}\,\,\,{#2}}\atop{{#3}}};
                     {#4},{#5}\right)}
\nc{\bra}[1]{\langle #1 |}
\nc{\ket}[1]{| #1 \rangle}
\nc{\qp}[2]{({#1}\, ; \, {#2})_{\infty}}
\nc{\qpf}[1]{({#1}\, ; \, q^4)_{\infty}}
\nc{\pp}[1]{({#1}\, ; \, p)_{\infty}}
\nc{\qpp}[1]{({#1}\, ; \, p, q^4)_{\infty}}
\nc{\sect}{\section}
\nc{\ssect}{\subsection}
\nc{\sssect}{\subsubsection}
\nc{\isomo}{\buildrel {\sim} \over \longrightarrow}
\nc{\Aff}{\operatorname{Aff}}
\nc{\ot}{\otimes}
\nc{\br}[1]{\begin{array}{#1}}
\nc{\er}{\end{array}}
\nc{\bev}[1]{\begin{equation}\begin{array}{#1}}
\nc{\eeq}{\end{equation}}
\nc{\be}{\begin{eqnarray}}
\nc{\ee}{\end{eqnarray}}
\nc{\ben}{\begin{eqnarray*}}
\nc{\een}{\end{eqnarray*}}
\nc{\bec}{\begin{equation}\begin{array}{lll}}
\nc{\eec}{\end{array}\end{equation}}
\nc{\ed}{\end{document}}
\nc{\half}{\ensuremath{\frac{1}{2}}}
\nc{\Hom}{\operatorname{Hom}}
\nc{\End}{\operatorname{End}}
\nc{\vac}{|\textrm{vac}\rangle}
\nc{\dvac}{\langle\textrm{vac}|}
\nc{\id}{\operatorname{id}}
\nc{\ra}{\rightarrow}  
\nc{\lra}{\longrightarrow}
\nc{\uqp}{U^{\prime}_q (\widehat{sl}_2)}
\nc{\vsl}{V(\sigma(\lambda))}
\nc{\vl}{V(\lambda)}  
\nc{\bu}{\bullet}
\nc{\an}{{\ell}}
\nc{\slth}{\widehat{\mathfrak{sl}}_2\hskip 1pt}
\nc{\uq}{U_q(\slth)}
\nc{\ws}{\;\;}
\nc{\qu}{{1\ov 4}}
\nc{\hif}{\hb{ if }}
\nc{\hev}{\hb{ is even }}
\nc{\hod}{\hb{ is odd }}
\nc{\Tr}{{\rm Tr}}
\nc{\hb}{\hbox}
\nc{\nn}{\nonumber} 
\nc{\curlra}{\buildrel{\sim}\over\longrightarrow}
\nc{\epp}{\varepsilon^{\prime}} 
\nc{\ol}{\overline}
\nc{\pl}{\prod\limits} 
\nc{\sli}{\sum\limits} 
\nc{\nin}{\noindent}

\nc{\ga}{\alpha}
\nc{\gb}{\beta}
\nc{\gd}{\delta}
\nc{\gep}{\varepsilon}
\nc{\gz}{\zeta}
\nc{\gt}{\theta}
\nc{\gk}{\kappa}
\nc{\gl}{\lambda}
\nc{\gp}{\phi}
\nc{\gs}{\sigma}
\nc{\go}{\omega}
\nc{\gn}{\nu}
\nc{\gr}{\rho}

\nc{\s}{\sigma}
\nc{\ep}{\varepsilon}
\nc{\z}{\zeta}
\nc{\g}{\gamma}
\nc{\zi}{\zeta^{-1}}

\nc{\gG}{\Gamma}
\nc{\gD}{\Delta}
\nc{\gT}{\Theta}
\nc{\gL}{\Lambda}
\nc{\gO}{\Omega}
\nc{\gP}{\Phi}

\nc{\cL}{\mathcal{L}}
\nc{\cF}{\mathcal{F}}
\nc{\cA}{\mathcal{A}}
\nc{\cP}{\mathcal{P}}
\nc{\cR}{\mathcal{R}}
\nc{\cS}{\mathcal{S}}
\nc{\cN}{\mathcal{N}}
\nc{\cD}{\mathcal{D}}
\nc{\cH}{\mathcal{H}}
\nc{\cO}{\mathcal{O}}
\nc{\cT}{\mathcal{T}}

\nc{\C}{\mathbb{C}}
\nc{\Q}{\mathbb{Q}}
\nc{\R}{\mathbb{R}}
\nc{\Z}{\mathbb{Z}}

\nc{\fg}{\mathfrak{g}}

\nc{\bi}{\bar{i}}
\nc{\bj}{\bar{j}}
\nc{\bp}{\bar{p}}
\nc{\bR}{\bar{R}}
\nc{\bgr}{\bar{\rho}}
\nc{\bA}{\bar{\alpha}}
\nc{\bB}{\bar{\beta}}
\nc{\bC}{\bar{\gamma}}
\nc{\by}{\bar{y}}

\nc{\tf}{\tilde{f}}
\nc{\te}{\tilde{e}}
\nc{\ts}{\tilde{s}}
\nc{\tgP}{\widetilde{\Phi}}
\nc{\tgPs}{\tilde{\Psi}}
\nc{\tgn}{\tilde{\nu}}
\nc{\tgl}{\tilde{\lambda}}
\nc{\txi}{\tilde{\xi}}

\nc{\cC}{\check{c}}
\nc{\cB}{\check{b}}

\nc{\goto}{\mapsto}
\nc{\embed}{\hookrightarrow}
\nc{\rien}{\emptyset}
\nc{\lb}[1]{\label{#1}}
\nc{\Nt}{\frac{N}{2}}
\nc{\vn}{\hspace*{-33truemm}}
\nc{\vm}{\hspace*{-0truemm}}

\begin{document}
\begin{flushright}
HWM00-6\\[10mm]
\end{flushright}
\begin{center}
{\LARGE \bf Impurity Operators in RSOS Models\\[10mm] }
{\large \bf Robert Weston}\\[6mm]
{\it Department of Mathematics, Heriot-Watt University,\\
Edinburgh EH14 4AS, UK.\\ R.A.Weston@ma.hw.ac.uk}\\[5mm]
March 2000\\[10mm]
\end{center}
\begin{abstract}
\noindent 
We give a construction of impurity operators in the `algebraic
analysis' picture of RSOS models. Physically, these operators
are half-infinite insertions of certain fusion-RSOS Boltzmann weights. They
are the face analogue of insertions of higher spin lines in vertex
models. Mathematically, they are given in terms of intertwiners of
$\uq$ modules. We present a detailed perturbation theory check of the
conjectural correspondence between the physical and mathematical
constructions for a particular simple example.
\end{abstract}
\nopagebreak
\setcounter{equation}{0}
\section{Introduction}
The `algebraic analysis' approach to solvable lattice models was
developed by the Kyoto group in the 1990s \cite{JM}. The key
feature of this approach is to identify the half-infinite space on
which the corner transfer matrix acts with an infinite-dimensional
module of the underlying non-Abelian symmetry algebra of the lattice
model. The simplest example is the anti-ferromagnetic 6-vertex model,
in which the half-infinite space is identified with $V(\gL_i)$, a 
level-one highest-weight module of the algebra $\uq$ \cite{Daval,JM}. 
The choice of the subscript $i\in\{0,1\}$ corresponds to the choice of
one of the two possible anti-ferromagnetic boundary conditions. A
somewhat more complicated example is that of RSOS models
\cite{JMO93}. Here, the half-infinite space is identified with the
space $\Omega_{\xi,\eta;\gl}$ that occurs in the decomposition of the
tensor product of $\uq$ highest-weight modules
\ben V(\xi)\ot V(\eta) \simeq \bigoplus_{\gl} V(\gl)\ot 
\gO_{\xi,\eta;\gl}\,,\label{iso1}\een
where $\xi$ and $\eta$ are level $k-n$ and level $n$ dominant integral
weights, and the sum is over all level $k$ dominant integral weights
(see \cite{JMO93} and below for more details).
Again, the choice of $\xi$, $\eta$ and $\gl$ in
$\Omega_{\xi,\eta;\gl}$ corresponds to the choice of
boundary conditions for the lattice model.

The other main step in the algebraic analysis approach is to 
identify the half-infinite transfer matrices of the lattice models
with certain intertwiners, or vertex operators, of the symmetry
algebra. For the 6-vertex model, the half-infinite transfer matrix
is identified with the $\uq$ intertwiner
\ben \gP_{\gL_i}^{\gL_{1-i} V^{(1)}}(\z): 
V(\gL_i) \rightarrow V(\gL_{1-i})\ot V^{(1)}_{\z},\een
where $V^{(1)}_\z$ is a spin-$\half$ $\uq$ evaluation module.
For the RSOS case, the 
situation is again slightly more complicated. If $\gl$ and $\gl'$ 
are level-$k$ dominant integral weights, then the $\uq$ intertwiner 
\ben \gP^{\gl' V^{(n)}}_{\gl}(\z): 
V(\gl)\rightarrow V(\gl')\ot V^{(n)}_\z\een
exists if and only if  $1\leq n\leq k$
and the pair $(\gl,\gl')$ is `$n$-admissible' as defined by \mref{adm} below.
Consider the intertwiner
\ben V(\xi) \ot V(\eta)
    \xrightarrow{1\ot \Phi_{\eta}^{\gs(\eta) V^{(n)}}(\z)}
    V(\xi) \ot  V(\gs(\eta)) \ot V^{(n)}_{\z},\een
where $\gs(a\gL_1+b\gL_0)=(b\gL_1+a\gL_0).$ 
Under the isomorphism \mref{iso1}, we can identify
\ben 1\ot \Phi_{\eta}^{\gs(\eta) V^{(n)}} = 
         \bigoplus_{(\gl,\gl')} \gP^{\gl' V^{(n)}}_{\gl}(\z)
           \ot X_{\gl}^{\gl'}(\z),\een
where the sum is over all $n$-admissible pairs $(\gl,\gl')$ of 
level-$k$ dominant integral weights. This identification defines the
operator
\ben X_{\gl}^{\gl'}(\z): \Omega_{\xi,\eta;\gl}
\rightarrow \Omega_{\xi,\gs(\eta);\gl'}\een (see (\ref{cdef1})). It is  $X_{\gl}^{\gl'}(\z)$
that is identified with
the half-infinite transfer matrix of the RSOS lattice model
\cite{JMO93}.

The role of impurity operators in the 6-vertex model was considered in
\cite{Nak96,MW97}. 
For vertex models, the term impurity operator refers to
the half-infinite transfer matrix corresponding to the insertion of a spin-$\frac{n}{2}$
line into a spin-$\half$ 6-vertex model. In \cite{MW97}, this 
operator was identified with the $\uq$
intertwiner
\ben \gP^{(n-1,n)}_{\gL_i}: 
V^{(n-1)}_{\z}\ot V(\gL_i) \rightarrow V(\gL_{1-i})\ot
V^{(n)}_{\z}\een
(which exists for all $n>1$).

In this paper, we shall consider analogous impurity operators in
RSOS models. The physical impurity operator corresponds to
the half-infinite
insertion of $W_k^{(m,n)}$ RSOS weights 
(see Section \ref{lattsec}) into a lattice made up of $W_k^{(n,n)}$ 
weights. The mathematical object with which this lattice
operator will be identified is defined in terms of the composition
\ben V(\xi) \ot V(\eta)
    &&\xrightarrow{ \Phi_{\xi}^{\xi' V^{(m-n)}}(\z)}
    V(\xi') \ot V^{(m-n)}_{\z} \ot V(\eta)\\&&
    \xrightarrow{\Phi^{(m-n,m)}_{\eta}(\z)}
    V(\xi') \ot  V(\s(\eta)) \ot V^{(m)}_{\z},\een
where $k\geq m>n\geq 1$ (see (\ref{int2}) for a definition of $\Phi^{(m-n,m)}_{\eta}(\z)$). Under the isomorphism \mref{iso1}, we 
identify
\ben  \Phi_{\eta}^{(m-n,m)}(\z)\circ \gP_{\xi}^{\xi' V^{(m-n)}}(\z) = 
         \bigoplus_{(\gl,\gl')} \gP^{\gl' V^{(m)}}_{\gl}(\z)
           \ot Z_{\xi\,\gl;m}^{\xi'\gl'}(\z),\een
where again the sum is over all all $n$-admissible pairs $(\gl,\gl')$.
This equality defines the operator 
\ben  Z_{\xi\,\gl;m}^{\xi'\gl'}(\z):
\gO_{\xi,\eta;\gl}\rightarrow \Omega_{\xi',\gs(\eta);\gl'}.\een
It is this that we shall identify 
with the RSOS impurity operator (a statement of the conjectural
identification
is given in (\ref{conj})).

The plan of this paper is as follows: in Section 2, we define the
necessary $\uq$ intertwiners and give some of their properties. In
Section 3, we recall some of the details of the algebraic analysis picture of RSOS
models and give our precise conjecture about the
realisation of impurity operators in this picture. We 
give the details of a perturbation theory check of this conjecture in
Section 4. We present 
a brief discussion of some possible future avenues of research opened by
this
work in Section 5.
In Appendix A, we
give the solution of the q-KZ equation for certain matrix elements of
intertwiners and use this solution in order to compute their 
commutation relations. In Appendix B, we give a proof of the
commutation relations of another type of intertwiner. Finally, we
list some formulae for the perturbative action of our different
intertwiners in Appendix C.

\setcounter{equation}{0}
\section{Properties of $\uq$ Intertwiners}
\ssect{Preliminaries}\label{defsec}
In this section, we shall define the $\uq$ intertwiners we need in
our discussion, and give some of their properties. Let us first recall
a few details about $\uq$. (See, for example, \cite{idzal93} for a fuller
account  - the only significant difference with our notation 
is that we use a  different evaluation module. Note also, that
although we use the notation $\uq$, we are actually referring to the
subalgebra generated by $e_i,f_i,t_i$ $(i=0,1)$.)
A weight
lattice $P_+=\Z_{\geq 0} \gL_0 \oplus \Z_{\geq 0} \gL_1$ occurs in the
definition of $\uq$. 
Let $h_0$ and $h_1$ denote the basis vectors for the lattice dual to
$P_+$, with $\langle h_i, \gL_j\rangle = \gd_{i,j}$.
Define the level $k\in \Z_{\geq 0}$
weight $\gl^{(k)}_a\in P_+$ by
\ben
\gl^{(k)}_a =  a \gL_1+ (k-a) \gL_0, \quad a\in \{0,1,\cdots,k\}.\een
Let $P_k^0$ be the set of such weights, i.e., 
\ben P_k^0=\{\gl_a^{(k)}\, |\, a\in\{0,1,\cdots,k\}\},\een
and define the function $\gs:P_+ \rightarrow P_+$ by
\ben \gs(a\gL_1+b\gL_0)= b\gL_1+a \gL_0.\een
We shall also use the notation 
\ben p=q^{2(k+2)},\quad s=\frac{1}{2(k+2)},\quad
\bar{\rho}=(\gL_1-\gL_0).\een

Suppose we choose an integer $N$ such that $k\geq N \geq 0$.
Then a pair of weights $(\gl^{(k)}_a,\gl^{(k)}_b)$ is said to be
`N-admissible' if 
\bec &(i) \quad a-b&\in \{N,N-2,\cdots, -N\},\\ 
    &(ii) \quad a+b&\in \{2k-N,2k-N-1,\cdots,N\}. \er\lb{adm}\end{equation}
In the case $N=0$, we have $a=b$. When $N=1$, the second condition follows from the first.
It is useful to introduce the notation $A^{(N)}_k$ for the
set of admissible pairs, i.e., 
\ben A^{(N)}_k = \{(\gl,\gl')\in P^0_k\times P^0_k\, | \, (\gl,\gl')
\ws\hbox{are $N$-admissible}\}.\een
Note that if $(\gl_1,\gl'_1)\in A_{k_1}^{(N_1)}$ and
$(\gl_2,\gl'_2)\in A_{k_2}^{(N_2)}$,
then it follows that $(\gl_1+\gl_2,\gl'_1+\gl'_2)\in A_{k_1+k_2}^{(N_1+N_2)}$.

We shall use two types of $\uq$ modules: irreducible highest weight 
modules $V(\gl)$ and evaluation modules $V^{(N)}_\z$. The irreducible
highest weight module $V(\gl)$ is generated by a highest weight vector 
$v_{\gl}$, defined by $e_i v_{\gl} =0$, $f_i^{\langle h_i,\gl\rangle
  +1}v_{\gl}=0$, for $i\in\{0,1\}$. We use the principally specialised 
spin-$\frac{N}{2}$ evaluation module $V_{\z}^{(N)}$ defined, in terms of weight
vectors $u^{(N)}_{i}$ $(i=0,1,\cdots, N)$, in Section 3.1 of \cite{MW97}.

We will also need the  \textit{R}-matrix, namely the $\uq$
intertwiner
\ben R^{(M,N)}(\z_1/\z_2):V^{(M)}_{\z_1}\ot V^{(N)}_{\z_2} \rightarrow
V^{(N)}_{\z_2}\ot V^{(M)}_{\z_1}.\een
The normalisation is fixed by
$R^{(M,N)}(\z)=\bR^{(M,N)}(\z)/\gk^{(M,N)}(\z)$,
where 
\be \bR^{(M,N)}(\z)(u^{(M)}_0\ot u^{(N)}_0)&=& (u^{(N)}_0\ot
u^{(M)}_0),\ws \hbox{and} \nn \\[2mm] \kappa^{(M,N)}(\z)&=&\z^{\min(M,N)}
 \frac{\qpf{q^{2+M+N}\z^2}\qpf{q^{2+|M-N|}\z^{-2}}}%
      {\qpf{q^{2+M+N}\z^{-2}} \qpf{q^{2+|M-N|}\z^2} }.\lb{kappa}\ee
(We use the standard notation $(a;b)_{\infty}=\pl_{n=0}^{\infty}
(1-a\, b^n)$.)
This is the normalisation that ensures crossing and unitarity for the 
\textit{R}-matrix - see \cite{HKMW98b} (this normalisation is also the one
that would give the vertex model with $R^{(M,N)}(\z)$ Boltzmann weights 
a partition function per site equal to one).
\ssect{Intertwiners}\label{intsec}
We shall make use of the following 
two types of $\uq$ intertwiner:
\be \gP_{\gl}^{\gl' V^{(N)}}(\z)&:& V(\gl)\rightarrow V(\gl')\ot
V^{(N)}_\z,\quad\ws (\gl,\gl')\in A^{(N)}_k,\quad N \in \{1,2,\cdots,k\},
\label{int1}\\[2mm]
\gP_{\gl}^{(N,N+k)}(\z)&:& V^{(N)}_\z \ot V(\gl)\rightarrow
V(\s(\gl))\ot  V^{(N+k)}_\z,\quad \ws\gl \in P^0_k, \quad N\in \Z_{>0}.\label{int2}\ee
It is shown in \cite{FR} that $\gP_{\gl}^{\gl' V^{(N)}}(\z)$ exists
and is unique up to a normalisation if and only if $(\gl,\gl')$ is an
$N$-admissible pair. The existence and uniqueness 
of $\gP_{\gl}^{(N,N+k)}(\z)$ is shown in \cite{HKMW98b}
(the $k=1$ operator was first introduced by Nakayashiki in \cite{Nak96}).
We fix the normalisation of $\gP_{\gl}^{\gl' V^{(N)}}(\z)$ by the requirement
\be
\gP_{\gl}^{\gl' V^{(N)}}(\z)&:& v_{\gl}\longmapsto v_{\gl'}\ot u^{(N)}_j+\cdots, 
\quad \hb{where}\ws \gl=\gl'+(N-2j)\bgr.
\label{vonorm}\ee
Here, \, $\cdots$ means terms involving $F v_{\gl'}$, where $F$ is 
some product of $f_0$ and $f_1$ generators.
The normalisation of $\gP_{\gl}^{(N,N+k)}(\z)$ is that given in
Section 5 of \cite{HKMW98b}.

Now, we shall give the commutation relations of the two types of
intertwiner \mref{int1} and \mref{int2}.
In \cite{FR}, Frenkel and Reshetikhin showed that the commutation relations
of \mref{int1} take the form
\be
R^{(M,N)}(\z) \gP^{\nu V^{(M)}}_{\mu}(\z_1) \gP^{\mu V^{(N)}}_{\gl}(\z_2)
= \sum_{\mu'}  \gP^{\nu V^{(N)}}_{\mu'}(\z_2) \gP^{\mu' V^{(M)}}_{\gl}(\z_1)
\cet{\gl}{\mu}{\mu'}{\nu}{\z}{(N,M)}{k}.\label{comm1}\ee
where $\z=\z_1/\z_2$,\, the sum is over $\{\mu'\in P^0_k\,|\, (\nu,\mu')\in A^{(N)}_k,
(\mu',\gl)\in A^{(M)}_k\}$,\,  and 
the connection coefficients $C^{(N,M)}_k$ satisfy the 
Yang-Baxter equation in its face formulation. 
As a special case, we have 
\ben \cet{\gl}{\mu}{\mu'}{\nu}{\z}{(k,k)}{k}= \gd_{\gl,\nu}
\gd_{\mu,\mu'} \gd_{\mu,\gs(\nu)}\een
(see \cite{idzal93}). In Appendix A, we 
solve the q-KZ equation to obtain the explicit
form \mref{cn1}--\mref{cnf} of the coefficients $C^{(N,1)}_k$ and $C^{(1,N)}_k$. In Appendix B, we
prove that the commutation relations of \mref{int2} are given by
\be
 R^{(N+k,N+k)}(\z)
  \gP^{(N,N+k)}_{\gs(\gl)}(\z_1)\gP^{(N,N+k)}_{\gl}(\z_2)= 
  \gP^{(N,N+k)}_{\s(\gl)}(\z_2) \gP^{(N,N+k)}_{\gl}(\z_1)
  R^{(N,N)}(\z)\label{comm2}.\ee
\ssect{Operators on the space $\Omega_{\xi,\eta;\gl}$}
Fix $\xi\in P^0_{k-n}$ and  $\eta\in P^0_{n}$ with $k > n\geq 1$.
Following \cite{JMO93} and \cite{JS97}, we consider the
decomposition 
\be V(\xi)\ot V(\eta) \simeq \bigoplus_{\gl\in P^0_k} V(\gl)\ot 
\gO_{\xi,\eta;\gl}.\label{iso2}\ee
Here $\gO_{\xi,\eta;\gl}$ denotes the space of highest vectors
\ben \gO_{\xi,\eta;\gl}= \{ v\in V(\xi)\ot V(\eta)\,|\, e_i v=0,\, t_i v=q^{<h_i,\gl>}
  v\}.\een

The existence of this decomposition allows us to use
the intertwiners \mref{int1} and \mref{int2} in order to define
certain operators on $\Omega_{\xi,\eta;\gl}$. Namely, we define 
\be X^{\gl'}_{\gl}(\z)&:& 
\Omega_{\xi,\eta;\gl}(\z)
 \rightarrow \Omega_{\xi,\gs(\eta);\gl'}
\quad \hbox{for}\ws (\gl,\gl')\in A^{(n)}_k,\label{Xdef}\\[2mm]
Z^{\xi'\gl'}_{\xi\,\gl;m}(\z)&:&
\gO_{\xi,\eta;\gl}\rightarrow \Omega_{\xi',\gs(\eta);\gl'},
\quad \hbox{for}\ws (\gl,\gl')\in A^{(m)}_k,\
(\xi,\xi')\in A^{(m-n)}_{k-n}, \  k\geq m>n, \label{Zdef}\ee 
via the identifications
\be
  \Phi_{\eta}^{\gs(\eta) V^{(n)}}(\z) &=&
         \bigoplus_{(\gl,\gl')\in A^{(n)}_{k}} \gP^{\gl' V^{(n)}}_{\gl}(\z)
           \ot X_{\gl}^{\gl'}(\z),\label{cdef1}\\
  \Phi_{\eta}^{(m-n,m)}(\z)\circ \gP_{\xi}^{\xi' V^{(m-n)}}(\z)&=&
         \bigoplus_{(\gl,\gl')\in A^{(m)}_{k}} \gP^{\gl' V^{(m)}}_{\gl}(\z)
           \ot Z_{\xi\,\gl;m}^{\xi'\gl'}(\z).\label{cdef2}\ee
It should be clear from the subscripts on which part of $V(\xi)\ot
V(\eta)$ the operators on the left-hand side act on.
In Section \ref{impsec}, we shall use a single
notation for both \mref{Xdef} and \mref{Zdef}, by defining 
$Z_{\xi\,\gl;n}^{\xi\gl'}(\z)$ by
$Z_{\xi\,\gl;n}^{\xi\gl'}(\z)= X_{\gl}^{\gl'}(\z)$.

The commutation relations of $X$ and $Z$ follow from their
definitions \mref{cdef1} and \mref{cdef2}, and from
\mref{comm1} and \mref{comm2}. We find that, acting on 
$\gO_{\xi,\eta;\gl}$, we have
  \ben
  \sum_{\tgl\in P^0_k} 
  &&\cet{\gl}{\tgl}{\gl'}{\gl''}{\z_1/\z_2}{(n,n)}{k}
  X_{\tgl}^{\gl''}(\z_1)
  X_{\gl}^{\tgl}(\z_2) =
  X_{\gl'}^{\gl''}(\z_2)
  X_{\gl}^{\gl'}(\z_1),
  \nn\\[3mm]
&&\sum_{\tgl\in P^0_k} \cet{\gl}{\tgl}{\gl'}{\gl''}{\z_1/\z_2}{(m,m)}{k}
  Z_{\xi'\,\tgl;m}^{\xi''\gl''}(\z_1)
  Z_{\xi\,\gl;m}^{\xi\tgl}(\z_2) = \\
&& \sum_{\txi\in P^0_{k-n}} Z_{\txi\,\gl';m}^{\xi''\gl''}(\z_2)
  Z_{\xi\,\gl;m}^{\txi\gl'}(\z_1) 
\cet{\xi}{\xi'}{\txi}{\xi''}{\z_1/\z_2}{(m-n,m-n)}{k-n}.
\een

\setcounter{equation}{0}
\section{The Algebraic Analysis Picture of RSOS models}
\ssect{The RSOS lattice model}\lb{lattsec}
Let us define lattice Boltzmann weights $W^{(m,n)}_k$ with
$k\geq m,n \geq 1$ by
\ben \wet{\gl}{\mu}{\mu'}{\nu}{\z}{(m,n)}{k}=
\cet{\nu}{\mu}{\mu'}{\gl}{\z}{(n,m)}{k},\een
where $(\gl,\mu),\  (\mu',\nu) \in A^{(m)}_k$ and 
$ (\gl,\mu'),\ (\mu,\nu) \in A^{(n)}_k$, and where the connection
coefficients $C^{(n,m)}_k$ are defined via \mref{comm1}. Then, it
follows from \mref{comm1} and from the Yang-Baxter equation and unitarity property of
$R^{(m,n)}(\z)$ (see \cite{HKMW98b}) that $W^{(m,n)}_k$ has the 
analogous face properties:
\ben
&&\sli_{\nu\in P^0_k} 
\wet{\ga}{\nu}{\mu}{\gl}{\z_2/\z_3}{(n,\ell)}{k}
\wet{\ga}{\gb}{\nu}{\g}{\z_1/\z_2}{(m,n)}{k}
\wet{\nu}{\g}{\gl}{\gd}{\z_1/\z_3}{(m,\ell)}{k}\\[2mm]
&&= \sli_{\nu\in P^0_k} 
\wet{\ga}{\gb}{\mu}{\nu}{\z_1/\z_3}{(m,\ell)}{k}
\wet{\mu}{\nu}{\gl}{\gd}{\z_1/\z_2}{(m,n)}{k}
\wet{\gb}{\g}{\nu}{\gd}{\z_2/\z_3}{(n,\ell)}{k},\\[3mm]
&&\sli_{\mu'\in P^0_k} 
\wet{\gl}{\mu}{\mu'}{\nu}{\z}{(m,n)}{k}
\wet{\gl}{\mu'}{\ga}{\nu}{\z^{-1}}{(n,m)}{k}= \gd_{\mu,\ga}.
\een

We can prove some additional
properties of $W^{(n,1)}_k$ and $W^{(1,n)}_k$  by making use of the explicit formulae for these
weights given in Appendix A. The first property relates $W^{(n,1)}_k$
and $W^{(1,n)}_k$:
\be 
\wet{\gl}{\mu}{\mu'}{\nu}{\z}{(n,1)}{k}=\wet{\nu}{\mu}{\mu'}{\gl}{\z}{(1,n)}{k}.\lb{prop1}\ee
The second property is that of crossing symmetry:
\be \wet{\gl}{\mu}{\mu'}{\nu}{-q^{-1}\z}{(n,1)}{k} =
    \frac{G(\gl,\mu')}{G(\mu,\nu)} 
    \wet{\mu'}{\gl}{\nu}{\mu}{\z^{-1}}{(1,n)}{k},\lb{prop2}\ee
where
\ben 
G(\gl_a^{(k)},\gl_{a+1}^{(k)})= \frac{\gG_p\big(1-2s(a+1)\big)}{\gG_p\big(1-2s(a+2)\big)}, \quad
G(\gl_a^{(k)},\gl_{a-1}^{(k)})= \frac{\gG_p\big(2s(a+1)\big)}{\gG_p\big(2sa\big)}.\een
Here, $\gG_p$ is the `q-gamma' function defined in \mref{fndef}, and 
$p$ and $s$ are as defined in Section \ref{defsec}.
We anticipate that formulae similar to \mref{prop1} and \mref{prop2}
will hold for the general $W^{(m,n)}_k$.

We shall define our lattice model by associating a Boltzmann weight 
$W^{(m,n)}_k$ with a configuration of $P_k^{0}$ weights around a face in the
following way: \newline
$ \hspace*{20mm}\wet{\gl}{\mu}{\mu'}{\nu}{\z}{(m,n)}{k}\quad  \sim $
\setlength{\unitlength}{0.0003in}
\begingroup\makeatletter\ifx\SetFigFont\undefined%
\gdef\SetFigFont#1#2#3#4#5{%
  \reset@font\fontsize{#1}{#2pt}%
  \fontfamily{#3}\fontseries{#4}\fontshape{#5}%
  \selectfont}%
\fi\endgroup%
{\renewcommand{\dashlinestretch}{30}
\begin{picture}(3317,3276)(0,1500)
\path(525,2928)(525,528)
\path(525,2928)(2925,2928)
\path(2925,2928)(2925,528)
\path(525,528)(2925,528)
\path(825,2928)(525,2628)
\put(3225,3153){\makebox(0,0)[lb]{\smash{{{\SetFigFont{12}{14.4}{\rmdefault}{\mddefault}{\updefault}$\mu$}}}}}
\put(3000,3){\makebox(0,0)[lb]{\smash{{{\SetFigFont{12}{14.4}{\rmdefault}{\mddefault}{\updefault} }}}}}
\put(3075,0){\makebox(0,0)[lb]{\smash{{{\SetFigFont{12}{14.4}{\rmdefault}{\mddefault}{\updefault}  $\nu$}}}}}
\put(0,3078){\makebox(0,0)[lb]{\smash{{{\SetFigFont{12}{14.4}{\rmdefault}{\mddefault}{\updefault}  }}}}}
\put(150,3153){\makebox(0,0)[lb]{\smash{{{\SetFigFont{12}{14.4}{\rmdefault}{\mddefault}{\updefault}$\gl$}}}}}
\put(150,0){\makebox(0,0)[lb]{\smash{{{\SetFigFont{12}{14.4}{\rmdefault}{\mddefault}{\updefault}$\mu'$}}}}}
\put(1500,1500){\makebox(0,0)[lb]{\smash{{{\SetFigFont{12}{14.4}{\rmdefault}{\mddefault}{\updefault}$\zeta$}}}}}
\end{picture}
}

\vspace*{10mm}

\nin Here, one corner is marked in order to give an orientation to the diagram.

The partition function of our lattice model will be a weighted sum
over the configurations of the weights at the corners of faces. 
In order to specify this partition function in the infinite-volume
limit, we must specify the boundary conditions for these
configurations at large distances from the centre of the lattice.
We will choose these boundary configurations such that the associated 
Boltzmann weights are maximal. Let us now fix $q$ and $\z$ such that
$0<-q<\z^{-1}<1$. Then from the explicit formula \mref{cn1} we find that
the largest Boltzmann weights $W^{(n,1)}_k$, $n\geq 1$, are those of the form
\be \wet{\xi'+\gL_i}{\xi+\gL_{1-i}}{\xi'+\gL_{1-i}}{\xi+\gL_i}{\z}{(n,1)}{k},\quad
{\rm with  }\ws (\xi,\xi') \in A^{(n-1)}_{k-1}.\label{adm1}\ee
We assume, by extension, that when $k\geq m\geq n\geq 1$, the largest weights
are those of the form
\be \wet{\xi'+\eta}{\xi+\s(\eta)}{\xi'+\s(\eta)}{\xi+\eta}{\z}{(m,n)}{k},\quad
{\rm with  }\ws (\xi,\xi') \in A^{(m-n)}_{k-n},\ws \eta\in P^0_{n}.
\label{adm2}\ee

Now, following the approach to RSOS models
described in \cite{JMO93} and \cite{Fodal93}, 
we consider the $(n,n)$ RSOS lattice model, that is, the RSOS model
constructed in 
terms of $W^{(n,n)}_k$, $n\geq 1$ weights. The boundary conditions
will
be labelled by a pair of weights $(\xi,\eta)\in P^0_{k-n}\times
P^0_{n}$ in the following way: if the position of the central weight is labelled 1, 
then we consider weight configurations such that beyond a large but finite
numbers of sites out from the centre, the weights at odd positions (along the vertical or
horizontal directions) are fixed to be $\xi+\eta$, and the weights 
at even positions are fixed to be $\xi+\s(\eta)$.

The North-West corner transfer matrix $A_{NW}^{(\xi,\eta;\gl)}(\z)$
with this boundary condition, and 
with the centre weight fixed to $\gl\in P^0_k$,
is represented graphically by

\vspace*{5mm}
\setlength{\unitlength}{0.00058333in}
\begingroup\makeatletter\ifx\SetFigFont\undefined%
\gdef\SetFigFont#1#2#3#4#5{%
  \reset@font\fontsize{#1}{#2pt}%
  \fontfamily{#3}\fontseries{#4}\fontshape{#5}%
  \selectfont}%
\fi\endgroup%
{\renewcommand{\dashlinestretch}{30}
\begin{picture}(7659,4194)(0,-10)
\path(5400,3933)(4500,3933)
\path(4500,3933)(4500,3033)
\path(4500,3033)(3600,3033)
\path(3600,3033)(3600,2133)
\path(5400,3033)(4425,3033)
\path(5400,2133)(3600,2133)
\path(3600,2133)(2700,2133)
\path(2700,2133)(2700,1233)
\path(2700,1233)(5400,1233)
\path(1800,1233)(2850,1233)
\path(1800,1233)(1800,333)
\path(1800,333)(5400,333)
\path(5400,3933)(5400,333)
\path(4500,3033)(4500,333)
\path(3600,2133)(3600,333)
\path(2700,1233)(2700,333)
\path(4650,3933)(4500,3783)
\path(3750,3033)(3600,2883)
\path(2850,2133)(2700,1983)
\path(1950,1233)(1800,1083)
\path(2850,1233)(2700,1083)
\path(3750,2133)(3600,1983)
\path(4650,3033)(4500,2883)
\path(4650,2133)(4500,1983)
\path(4650,1233)(4500,1083)
\path(3750,1233)(3600,1083)
\put(5475,33){\makebox(0,0)[lb]{\smash{{{\SetFigFont{11}{13.2}{\rmdefault}{\mddefault}{\updefault}$\gl$}}}}}
\put(5475,3933){\makebox(0,0)[lb]{\smash{{{\SetFigFont{11}{13.2}{\rmdefault}{\mddefault}{\updefault} }}}}}
\put(5475,4083){\makebox(0,0)[lb]{\smash{{{\SetFigFont{11}{13.2}{\rmdefault}{\mddefault}{\updefault}$\ga$}}}}}
\put(4125,4008){\makebox(0,0)[lb]{\smash{{{\SetFigFont{11}{13.2}{\rmdefault}{\mddefault}{\updefault}$\gb$}}}}}
\put(4125,3108){\makebox(0,0)[lb]{\smash{{{\SetFigFont{11}{13.2}{\rmdefault}{\mddefault}{\updefault}$\ga$}}}}}
\put(3225,2958){\makebox(0,0)[lb]{\smash{{{\SetFigFont{11}{13.2}{\rmdefault}{\mddefault}{\updefault}$\gb$}}}}}
\put(3225,2208){\makebox(0,0)[lb]{\smash{{{\SetFigFont{11}{13.2}{\rmdefault}{\mddefault}{\updefault}$\ga$}}}}}
\put(2325,2208){\makebox(0,0)[lb]{\smash{{{\SetFigFont{11}{13.2}{\rmdefault}{\mddefault}{\updefault}$\gb$}}}}}
\put(2325,1383){\makebox(0,0)[lb]{\smash{{{\SetFigFont{11}{13.2}{\rmdefault}{\mddefault}{\updefault}$\ga$}}}}}
\put(1350,1233){\makebox(0,0)[lb]{\smash{{{\SetFigFont{11}{13.2}{\rmdefault}{\mddefault}{\updefault}$\gb$}}}}}
\put(1350,33){\makebox(0,0)[lb]{\smash{{{\SetFigFont{11}{13.2}{\rmdefault}{\mddefault}{\updefault}$\ga$}}}}}
\put(4875,708){\makebox(0,0)[lb]{\smash{{{\SetFigFont{11}{13.2}{\rmdefault}{\mddefault}{\updefault}$\z$}}}}}
\put(4875,1608){\makebox(0,0)[lb]{\smash{{{\SetFigFont{11}{13.2}{\rmdefault}{\mddefault}{\updefault}$\z$}}}}}
\put(4875,2508){\makebox(0,0)[lb]{\smash{{{\SetFigFont{11}{13.2}{\rmdefault}{\mddefault}{\updefault}$\z$}}}}}
\put(4875,3408){\makebox(0,0)[lb]{\smash{{{\SetFigFont{11}{13.2}{\rmdefault}{\mddefault}{\updefault}$\z$}}}}}
\put(3975,2508){\makebox(0,0)[lb]{\smash{{{\SetFigFont{11}{13.2}{\rmdefault}{\mddefault}{\updefault}$\z$}}}}}
\put(3975,1608){\makebox(0,0)[lb]{\smash{{{\SetFigFont{11}{13.2}{\rmdefault}{\mddefault}{\updefault}$\z$}}}}}
\put(3975,708){\makebox(0,0)[lb]{\smash{{{\SetFigFont{11}{13.2}{\rmdefault}{\mddefault}{\updefault}$\z$}}}}}
\put(3075,1608){\makebox(0,0)[lb]{\smash{{{\SetFigFont{11}{13.2}{\rmdefault}{\mddefault}{\updefault}$\z$}}}}}
\put(3075,708){\makebox(0,0)[lb]{\smash{{{\SetFigFont{11}{13.2}{\rmdefault}{\mddefault}{\updefault}$\z$}}}}}
\put(2175,708){\makebox(0,0)[lb]{\smash{{{\SetFigFont{11}{13.2}{\rmdefault}{\mddefault}{\updefault}$\z$}}}}}
\put(0,2358){\makebox(0,0)[lb]{\smash{{{\SetFigFont{11}{13.2}{\rmdefault}{\mddefault}{\updefault}$A^{(\xi,\eta;\gl)}_{NW}(\z)=$}}}}}
\put(6525,2358){\makebox(0,0)[lb]{\smash{{{\SetFigFont{11}{13.2}{\rmdefault}{\mddefault}{\updefault}$\ga=\xi+\eta$,}}}}}
\put(7900,2358){\makebox(0,0)[lb]{\smash{{{\SetFigFont{11}{13.2}{\rmdefault}{\mddefault}{\updefault}$\gb=\xi+\s(\eta)$.}}}}}
\end{picture}
}

\vspace*{5mm}

 Let 
$\cH_{\xi,\eta;\gl}$ denote the space of eigenstates of 
$A_{NW}^{(\xi,\eta;\gl)}(\z)$ in the infinite volume
limit, such that $A_{NW}^{(\xi,\eta;\gl)}(\z):\cH_{\xi,\eta;\gl} \to
\cH_{\xi,\eta;\gl}$. Let $\ket{p}$ denote a restricted path
\ben \ket{p}=(\cdots,p(3),p(2),p(1)),\ws \hbox{with}\ws 
(p(\ell+1),p(\ell))\in A^{(n)}_k \ws \hbox{for}\ws \ell\geq 1.\een
Then, $\cH_{\xi,\eta;\gl}$ will be formally spanned by the path space
$\cP_{\xi,\eta;\gl}$ defined by
\ben
\cP_{\xi,\eta;\gl}=\{\ket{p}\,|\, p(\ell)=\xi+\s^{l-1}(\eta), \ws \ell\geq r> 1,\, p(1)=\gl\}.
\een
\ssect{The identification of $\gO_{\xi,\eta;\gl}$ and
  $\cH_{\xi,\eta;\gl}$}

Let us first introduce some extra notation. Define $\ket{p_{\xi,\eta}}$ to be
the `ground-state' path in $P_{\xi,\eta;\, \xi+\eta}$ given by
\ben \ket{p_{\xi,\eta}}= (\cdots p_{\xi,\eta} (3),
p_{\xi,\eta}(2),
p_{\xi,\eta} (1)), \ws \hbox{where}
\ws p_{\xi,\eta} (\ell)=\xi+\s^{\ell-1}(\eta).\een
Also, define  $v_{\xi,\eta}=v_{\xi}\ot v_{\eta}
\in \gO_{\xi,\eta;\,\xi+\eta}$.

A map $\iota:\gO_{\xi,\eta;\,\gl} \to \cH_{\xi,\eta;\,\gl}$ is given 
in \cite{JMO93}. 
In our notation, this map is given by 
\be
\iota(v)=\sli_{\ket{p}\in \cP_{\xi,\eta;\,\gl}} c(p,v) \ket{p},\lb{embed1}\ee
where
\be c(p,v)&=&\lim_{\ell\to \infty}
\frac{c^{\ell}(p,v)}{c^{\ell}( p_{\xi,\eta},v_{\xi,\eta})},\label{conv}\\[3mm]
c^{\ell}(p,v) &=& \bra{v_{\xi,\s^{\ell}(\eta)}}X^{p_{\xi,\eta}(\ell+1)}_{p(\ell)}(1)
\cdots X^{p(3)}_{p(2)}(1) X^{p(2)}_{\gl}(1)\ket{v}.\lb{embed3}\ee 

\nin It is a conjecture that \mref{conv} converges.
\ssect{The half-transfer matrix and impurity operators}\label{impsec}

First, we define the finite path space ${_N}\cP_{\xi,\eta;\gl}$ by
\be
{_N}\cP_{\xi,\eta;\gl}=\{(p(N+1),p(N),\cdots,p(1))\,|\,
(p(\ell+1),p(\ell))\in A^{(n)}_k,\, p(N+1)=\xi+\s^N(\eta),\, p(1)=\gl\}.\nn\ee
Let ${_N}\cH_{\xi,\eta;\gl}$ denote the vector space spanned by 
${_N}\cP_{\xi,\eta;\gl}$, and define $\rho_N$ to be the projection
operator $\rho_N:\cH_{\xi,\eta;\gl} \to {_N}\cH_{\xi,\eta;\gl}$.
Now we define the operator 
${_N}Z_{\xi\,\gl;m}^{\xi'\gl'}(\z)$ by
\ben
{_N}Z^{\xi'\gl'}_{\xi\,\gl;m}(\z)&:&
{_N}\cH_{\xi,\eta;\gl}\rightarrow {_N}\cH_{\xi',\gs(\eta);\gl'},
\quad \hbox{for}\ws (\gl,\gl')\in A^{(m)}_k,\
(\xi,\xi')\in A^{(m-n)}_{k-n}, \  k\geq m\geq n\geq 1, \nn\\[3mm]
{_N}Z^{\xi'\gl'}_{\xi\,\gl;m}(\z)\ket{p}&=& \sli_{\ket{p'}\in\, 
{_N}\cP_{\xi',\s(\eta);\gl'}} \pl_{\ell=1}^{N} 
\wet{p'(\ell+1)}{p(\ell+1)}{p'(\ell)}{p(\ell)}{\z}{(m,n)} \ws \ket{p' }.\een
Graphically, this operator is represented by
\vspace*{10mm}
\bec
\setlength{\unitlength}{0.00058333in}
\begingroup\makeatletter\ifx\SetFigFont\undefined%
\gdef\SetFigFont#1#2#3#4#5{%
  \reset@font\fontsize{#1}{#2pt}%
  \fontfamily{#3}\fontseries{#4}\fontshape{#5}%
  \selectfont}%
\fi\endgroup%
{\renewcommand{\dashlinestretch}{30}
\begin{picture}(3536,4464)(600,-10)
\path(2025,4092)(2925,4092)
\path(2025,4092)(2025,492)
\path(2025,492)(2925,492)
\path(2925,4092)(2925,492)
\path(2025,1392)(2925,1392)
\path(2025,2292)(2925,2292)
\path(2025,3192)(2925,3192)
\path(2175,4092)(2025,3942)
\path(2175,3192)(2025,3042)
\path(2175,2292)(2025,2142)
\path(2175,1392)(2025,1242)
\put(-500,2367){\makebox(0,0)[lb]{\smash{{{\SetFigFont{11}{13.2}{\rmdefault}{\mddefault}{\updefault}${_N}Z^{\xi'\gl'}_{\xi\,\gl;m}(\z)=$}}}}}
\put(2400,867){\makebox(0,0)[lb]{\smash{{{\SetFigFont{11}{13.2}{\rmdefault}{\mddefault}{\updefault}$\z$}}}}}
\put(2400,1767){\makebox(0,0)[lb]{\smash{{{\SetFigFont{11}{13.2}{\rmdefault}{\mddefault}{\updefault}$\z$}}}}}
\put(2400,2667){\makebox(0,0)[lb]{\smash{{{\SetFigFont{11}{13.2}{\rmdefault}{\mddefault}{\updefault}$\z$}}}}}
\put(2400,3567){\makebox(0,0)[lb]{\smash{{{\SetFigFont{11}{13.2}{\rmdefault}{\mddefault}{\updefault}$\z$}}}}}
\put(3075,4242){\makebox(0,0)[lb]{\smash{{{\SetFigFont{11}{13.2}{\rmdefault}{\mddefault}{\updefault}$\xi+\s^N(\eta)$}}}}}
\put(3075,42){\makebox(0,0)[lb]{\smash{{{\SetFigFont{11}{13.2}{\rmdefault}{\mddefault}{\updefault}$\gl$}}}}}
\put(3075,2000){\makebox(0,0)[lb]{\smash{{{\SetFigFont{11}{13.2}{\rmdefault}{\mddefault}{\updefault}$.$}}}}}
\put(3075,2250){\makebox(0,0)[lb]{\smash{{{\SetFigFont{11}{13.2}{\rmdefault}{\mddefault}{\updefault}$.$}}}}}
\put(3075,2500){\makebox(0,0)[lb]{\smash{{{\SetFigFont{11}{13.2}{\rmdefault}{\mddefault}{\updefault}$.$}}}}}
\put(1750,2000){\makebox(0,0)[lb]{\smash{{{\SetFigFont{11}{13.2}{\rmdefault}{\mddefault}{\updefault}$.$}}}}}
\put(1750,2250){\makebox(0,0)[lb]{\smash{{{\SetFigFont{11}{13.2}{\rmdefault}{\mddefault}{\updefault}$.$}}}}}
\put(1750,2500){\makebox(0,0)[lb]{\smash{{{\SetFigFont{11}{13.2}{\rmdefault}{\mddefault}{\updefault}$.$}}}}}
\put(3075,1317){\makebox(0,0)[lb]{\smash{{{\SetFigFont{11}{13.2}{\rmdefault}{\mddefault}{\updefault}$p(2)$}}}}}
\put(1125,117){\makebox(0,0)[lb]{\smash{{{\SetFigFont{11}{13.2}{\rmdefault}{\mddefault}{\updefault}$\gl'$}}}}}
\put(1125,1317){\makebox(0,0)[lb]{\smash{{{\SetFigFont{11}{13.2}{\rmdefault}{\mddefault}{\updefault}$p'(2)$}}}}}
\put(1125,3117){\makebox(0,0)[lb]{\smash{{{\SetFigFont{11}{13.2}{\rmdefault}{\mddefault}{\updefault}$p'(N)$}}}}}
\put(3075,3192){\makebox(0,0)[lb]{\smash{{{\SetFigFont{11}{13.2}{\rmdefault}{\mddefault}{\updefault}$p(N)$}}}}}
\put(1125,4242){\makebox(0,0)[lb]{\smash{{{\SetFigFont{11}{13.2}{\rmdefault}{\mddefault}{\updefault}$\xi'+\s^{N+1}(\eta)$}}}}}
\end{picture}
}
\lb{latop}\er\end{equation}

Let $\ket{v}\in \Omega_{\xi,\eta;\gl}$. Then 
our conjecture for the realisation of
${_N}Z^{\xi'\gl'}_{\xi\,\gl;m}(\z)$ in the algebraic analysis picture of
RSOS models is
\be
\lim_{N\to \infty}\, \frac{1}{f^{(m,n)}_N(\z,q)}\ {_N}Z^{\xi'\gl'}_{\xi\,\gl;m}(\z)
\circ \rho_N \circ \iota \ket{v}= \iota \circ Z^{\xi'\gl'}_{\xi\,\gl;m}(\z)
\ket{v},\label{conj}\ee
 where the function $f^{(m,n)}_N(\z,q)$ is a series in $q$, whose coefficients
are Laurent polynomials in $\z$ (this function may also depend
upon the values of $\xi,\eta,\gl,\xi'$ and
$\gl'$). $Z_{\xi\,\gl;m}^{\xi'\gl'}(\z)$ is defined by \mref{cdef1}
and \mref{cdef2} (with $Z_{\xi\,\gl;n}^{\xi\gl'}(\z)\equiv X_{\gl}^{\gl'}(\z)$).

When $m=n$, this conjecture gives us the algebraic analysis realisation
of the half-transfer matrix of our $(n,n)$ RSOS model. When $m>n$, it
gives us a realisation of the $(m,n)$ impurity operator, i.e., of the
operator made up from a half-infinite tower of $(m,n)$ weights
inserted into our $(n,n)$ RSOS model.

\setcounter{equation}{0}
\section{Perturbation Theory}\lb{pertsec}

In this section, we present the results of a perturbation 
theory check around $q=0$ of our conjecture \mref{conj}. We fix the values $(k,n)=(3,1)$ and 
check \mref{conj} for $m=1$ and for $m=2$.
$(k,n)=(3,1)$ is the simplest model for which both the half-transfer
 matrix and the $m=n+1$ impurity operator are non-trivial. 
The perturbation theory analysis involves three main steps. Step 1
is an extension of the analysis of the $k=2$ case carried out in \cite{JMO93}.
\vspace*{3mm}


\noindent {\bf Step 1}

\noindent First of all, we compute a perturbative expansion for $
\vac\in \cH^{2\gL_0,\gL_0;3\gL_0}$. This vector
is defined to be the
minimum eigenvalue eigenvector of the corner transfer matrix
 Hamiltonian $H_{\rm CTM}$.   $H_{\rm CTM}$ is in turn defined by
\ben H_{\rm CTM}= -\frac{dA_{NW}^{(2\gL_0,\gL_0;3\gL_0)}(\z)}{d\z}\Big|_{\z=1},\een
where $A_{NW}^{(2\gL_0,\gL_0;3\gL_0)}(\z)$ is the corner transfer
 matrix of the $(1,1)$ RSOS model with $k=3$.

We will use the following abbreviated notation for $(m,n)$ Boltzmann weights
\ben \wet{a}{b}{c}{d}{\z}{(m,n)}{k}\equiv
\wet{\gl_a^{(k)}}{\gl_b^{(k)}}{\gl_c^{(k)}}{\gl_d^{(k)}}{\z}{(m,n)}{k},\een
and we define $\overline{W}^{(1,1)}_k$ by
\ben \wet{a}{b}{c}{d}{\z}{(1,1)}{k}=
\frac{1}{\kappa^{(1,1)}(\z)}
\frac{\eta(\z^2)}{\eta(\z^{-2})}
\bwet{a}{b}{c}{d}{\z}{(1,1)}{k},\een
where $\kappa^{(1,1)}(\z)$ and $\eta(\z)$ are given by \mref{kappa} and
\mref{eta}.

Let us write out the weights for the (1,1) RSOS model (these come 
from formulae \mref{conn1}--\mref{conn2}). We have 
\be 
\bA_k(\z)\equiv \bwet{a}{a\pm 1}{a\pm 1}{a\pm2}{\z}{(1,1)}{k}&=&1,\label{w1}\\
\bB_k^{a\pm}(\z)\equiv\bwet{a}{a\pm 1}{a\mp 1}{a}{\z}{(1,1)}{k}&=&
q\frac{\gG_p(r_{\mp})\gG_p(r_{\mp})}{\gG_p(2s+r_{\mp})\gG_p(-2s+r_{\mp})}
\frac{\gT_p(\z^2)}{\gT(q^2\z^2)},
\label{w2}\\
\bC_k^{a\pm}(\z)\equiv\bwet{a}{a\pm 1}{a\pm 1}{a}{\z}{(1,1)}{k}&=&
\z \frac{\gT_p(q^2) \gT_p(p^{r_\pm}\z^2)}{\gT_p(q^2\z^2)
  \gT_p(p^{r_\pm})},
\label{w3}\ee
where $r_-=2(a+1)s$ and $r_+=1-r_-$. $\gG_p$ and $\gT_p$ are defined
in equation \mref{fndef}. The largest weight in our specified region
$0<-q<\z^{-1}<1$ is $\bC_k^{a\pm}(\z)$.

Noting that $\bA_k(1)=1$, $\bB_k^{a\pm}(1)=0$ and  $\bC_k^{a\pm}(1)=1$, our `renormalised'
corner transfer matrix Hamiltonian is given by
\be
H_{\rm CTM}^{r}=R-\sli_{\ell=1}^{\infty} \ell. O_{\ell}.\label{action}\ee
The operator $O_{\ell}$ acts as the identity on a path
 $\ket{p}\in \cP^{2\gL_0,\gL_0;3\gL_0}$ everywhere except on the
 triple $\big(p(\ell+2),p(\ell+1),p(\ell)\big)$, where its action is given 
 by
\ben
O_{\ell}(a\pm 2,a\pm 1, a)&=&0,\\
O_{\ell}(a,a\pm 1, a)&=& \cB^{a\pm}(a,a\mp 1,a)+\cC^{a\pm}(a,a\pm 1, a),\een
with
\ben \cB^{a\pm}\equiv \frac{d \bB_3^{a\pm}(\z)}{d\z}\Big|_{\z=1},\quad
\cC^{a\pm}\equiv \frac{d \bC_3^{a\pm}(\z)}{d\z}\Big|_{\z=1}.\een
Here, and elsewhere in this section, we use the abbreviated notation
$a$ to indicate the weight $\gl^{(k)}_a$.

Before giving the definition of the constant $R$, which fixes what we mean by
renormalised, let us introduce some
notation for certain paths $\ket{p}\in \cP_{2\gL_0,\gL_0;3\gL_0}$.
We use the notation $\ket{\rien}$ to indicate the ground-state path 
$\ket{p_{2\gL_0,\gL_0}}= (\cdots \cdots  1 \ws 0 \ws 1 \ws 0)$.
Then $\ket{2\ell+1}$, with $\ell>0$,  will indicate a path which differs
from $\ket{\rien}$ only in that $p(2\ell+1)=2$. Similarly,
$\ket{2\ell_1+1,2\ell_2+1,\cdots,2\ell_M }$ denotes a path that is the same as
$\ket{\rien}$ except that $p(2\ell_1+1)=2$, $p(2\ell_2+1)=2,\, \cdots, \
 p(2\ell_M+1)=2$. Finally, $\ket{2\ell+3,2\ell+2,2\ell+1}$ indicates
a path for which $p(2\ell+3)=2$, $p(2\ell+2)=3$ and $p(2\ell+1)=2$.
In Steps 2 and 3, we will use a very similar notation for paths
in other path spaces - but we will try to avoid confusion by always
specifying which path space we
are dealing with.

Now we come back to the meaning of \mref{action}.
$R=\sli_{\ell=1}^{\infty} \ell . R_{\ell}\id $\, is 
fixed by the requirements
\be H^{r}_{\rm CTM} \vac&=&0,\label{cc1}\\
    \langle{\rien}\vac &=&1.\label{cc2}\ee 
The $r$ superscript on $H_{CTM}^r$ indicates this choice of (re)normalisation.
The conditions \mref{cc1} and \mref{cc2} fix $R_{\ell}$ to be
\ben R_{2\ell-1}= \cC^{0+} , \quad 
R_{2\ell}=(\cC^{1-}+ \cB^{1+}\langle 2\ell+1\vac).\een

It remains only to solve $H^{r}_{\rm CTM} \vac=0$ perturbatively
by expanding both $H^{r}_{\rm CTM}$ and $\vac$ around $q=0$. We find
\be &&\vac=\ket{\rien}- q\sli_{\ell}\ket{2\ell+1} + 
q^2\Big(\sli_{\ell_1\gg \ell_2}\ket{2\ell_1+1,2\ell_2+1} +2 \sli_\ell
\ket{2\ell+3,2\ell+1}\Big)\nn\\
&&+ q^3\Big(
   2\sli_{\ell} \ket{2\ell+1} 
   -\sli_{\ell_1\gg \ell_2\gg \ell_3} \ket{2\ell_1+1,2\ell_2+1,2\ell_3+1}\nn\\
&&   -2\sli_{\ell_1\gg \ell_2+1}\ket{2\ell_1+1,2\ell_2+3,2\ell_2+1}
   -2\sli_{\ell_2\gg \ell_1}\ket{2\ell_2+3,2\ell_2+1,2\ell_1+1} \nn\\
&& -5 \sli_{\ell}\ket{2\ell+5,2\ell+3,2\ell+1} -\sli_{\ell} \ket{2\ell+3,2\ell+2,2\ell+1}
  \Big)+O(q^4),\lb{vac}\ee
where $\ell_1\gg \ell_2$  means $\ell_1>\ell_2+1$.
\vspace*{5mm}

\noindent{\bf Step 2}

\nin In this step, we will 
compute $\iota(\ket{v_{2\gL_0}\ot v_{\gL_0}})$, 
$\iota(X_{0}^{1}(\z)\ket{v_{2\gL_0}\ot v_{\gL_0}})$ and 
$\iota(Z_{00;2}^{12}(\z)\ket{v_{2\gL_0}\ot
  v_{\gL_0}})$ perturbatively. $X_{0}^{1}(\z)$ and
$Z_{00;2}^{12}(\z)$ are defined by \mref{cdef1} and
\mref{cdef2}, and
$\iota$ is defined by \mref{embed1}--\mref{embed3} (recall that
we are selectively indicating the weight $\gl^{(3)}_a$ by the integer $a$).

To find $\iota:\gO_{\xi,\eta,\gl}\to \cH_{\xi,\eta,\gl}$, 
we must calculate the perturbative action of 
$X_{\gl}^{\gl'}(\z):\gO_{\xi,\eta;\gl}\to \gO_{\xi,\s(\eta);\gl'}$.
To do this, it is useful if we make the identification 
\ben \gO_{\xi,\eta;\gl}=\Hom_{\uq}(V(\gl),V(\xi)\ot V (\eta)).\een
Then for $\ga\in \Hom_{\uq}(V(\gl),V(\xi)\ot V (\eta)) $,\, $X _{\gl}^{\gl'}(\z)(\ga)$ is defined
via the commutative diagram 

\vspace*{4mm}
\begin{center}
\begin{texdraw}
\fontsize{10}{13}\selectfont
\setunitscale 1
\drawdim mm
\arrowheadsize l:2.4 w:1.1 \arrowheadtype t:F
\textref h:C v:C
\move(0 0)
\bsegment
\htext(-1 0){$V(\gl)$}
\esegment
\move(0 20)
\bsegment
\htext(-0.5 0){$V(\xi)\ot V(\eta)$}
\esegment
\move(80 20)
\bsegment
\htext(-0.5 0){$V(\xi)\ot V(\s(\eta)\ot V^{(1)}_{\z}$}
\esegment
\move(80 0)
\bsegment
\htext(-0.5 0){$\sli_{\gl'} V(\gl')\ot V^{(1)}_\z$}
\esegment
\move(80 5)\avec(80 16)
\move(0 5)\avec(0 16)
\move(10 0)\avec(65 0)
\move(15 20)\avec(60 20)
\htext(40 25){$1\ot \Phi_{\eta}^{\s(\eta) V^{(1)}}(\z)$}
\htext(40 5){$\Phi_{\gl}^{\gl' V^{(1)}}(\z)$}
\htext(-5 10){$\ga$}
\htext(90 10){$X_{\gl}^{\gl'}(\z)(\ga)$}
\end{texdraw}
\end{center}
\vspace*{3mm}

Let us list the first few highest weight elements of the
various $V(\xi)\ot V(\eta)$ that we shall need in this section.
Note that if $w$ is such a highest weight element, then we have
the identification $w=\alpha(v_\gl)$.

\nin In $V(2\gL_0)\ot V(\gL_0)$, we have
 \ben
x_1^{(0)}&=& \vt \ot \vo,\\
x_1^{(2)}&=& \vt \ot f_0 \vo-q^2 \frac{1}{[2]} f_0\vt\ot \vo
,\\
x_2^{(2)}&=& \frac{1}{[2]}\vt \ot f_0 f_1 f_0 \vo 
- \frac{q^2}{[2]^2} f_0 \vt \ot f_1 f_0 \vo  
+\frac{q^4}{[2]^2} f_1 f_0 \vt\ot f_0 \vo\\[-1mm] &&
+q^6 \frac{1}{[2]^2([4]-[2])} (f_1 f_0^2 +(1-[3]) f_0f_1 f_0) \vt\ot \vo.
\een
\nin 
In $V(2\gL_0)\ot V(\gL_1)$:
\ben
x_1^{(1)}&=& \vt \ot \vw,\\
x_2^{(1)}&=& \frac{1}{[2]}\vt\ot f_0 f_1 \vw -\frac{q^2}{[2]} f_0 \vt \ot f_1 \vw
+\frac{q^4}{[2]^2} f_1 f_0 \vt \ot \vw,\\
x_1^{(3)}&=& \frac{1}{[2]} \vt\ot f_0^2 f_1 \vw -\frac{q^2}{[2]}f_0
\vt\ot f_0 f_1 \vw
+\frac{q^2}{[2]}
 f_0^2\vt\ot f_1 \vw \\[-1mm] && +\frac{q^6}{[2]([4]-[2])}
(f_0f_1f_0-f_1f_0^2)\vt\ot \vw.
\een
\nin In $V(\gL_1+\gL_0)\ot V(\gL_0)$:
\ben
&&\vn y_1^{(1)}= v_{\gL_1+\gL_0}\ot v_{\gL_0},\\
&&\vn y_1^{(3)}= v_{\gL_1+\gL_0}\ot f_0 v_{\gL_0} 
               -q f_0 v_{\gL_1+\gL_0}\ot v_{\gL_0},\\
&&\vn y_2^{(1)}=  \frac{1}{[2]}v_{\gL_1+\gL_0}\ot f_1 f_0 v_{\gL_0} 
                - q f_1 v_{\gL_1+\gL_0}\ot f_0 v_{\gL_0}\\[-2mm]
&&                +\frac{q^4}{1-[3]^2} (f_1f_0-[3]f_0 f_1) 
                v_{\gL_1+\gL_0}\ot v_{\gL_0}.\\
&&\vn y_2^{(3)}= \frac{1}{[2]} v_{\gL_1+\gL_0}\ot f_0 f_1 f_0
v_{\gL_0} -\frac{q}{[2]}f_0
v_{\gL_1+\gL_0}\ot f_1f_0 v_{\gL_0} \\[-2mm]
&&+ \frac{q^4}{1-[3]^2}(f_0f_1-[3]f_1f_0) v_{\gL_1+\gL_0}\ot f_0 v_{\gL_0}\\
&&-\frac{q^5}{1-[3]^2}(f_0^2f_1 -[3]f_0f_1f_0)v_{\gL_1+\gL_0}\ot v_{\gL_0}.\een
\nin And, in $V(\gL_1+\gL_0)\ot V(\gL_1):$
\ben
y_1^{(2)}=\overline{y_1^{(1)}}, \quad  y_1^{(0)}=\overline{y_1^{(3)}}, \quad
y_2^{(2)}= \overline{y_2^{(1)}}, \quad y_2^{(0)} = \overline{y_2^{(3)}},\een
where the bar operation exchanges $0$ and $1$ indices, e.g.,
 $\overline{y_1^{(3)}}=v_{\gL_1+\gL_0}\ot f_1 v_{\gL_1} -q f_1 v_{\gL_1+\gL_0}\ot v_{\gL_1}$.

\nin The notation is such that $x_{i}^{(a)}\in
\gO_{2\gL_0,\gL_j;\gl^{(3)}_a}$ and $y_{i}^{(a)}\in
\gO_{\gL_1+\gL_0,\gL_j;\gl^{(3)}_a}$ 
(with $j\in\{0,1\}$).

We can then calculate the perturbative action of $X_{\gl}^{\gl'}(\z)$
on these vectors by making use of the perturbative action of
$\Phi_{\eta}^{\s(\eta)V^{(1)}}\!(\z)$ and
$\Phi_{\gl}^{\gl' V^{(1)}}(\z)$ 
given in Appendix \ref{voapp}. We find 
\bev{ll}
 X_0^1(\z)(x_1^{(0)}) = x_1^{(1)} + \z^2 q^3 x_2^{(1)} +
\cdots,  &\\[2mm]
X_1^0(\z)(x_1^{(1)}) = x_1^{(0)} +  \cdots,   
&X_1^0(\z)(x_2^{(1)}) = \z^{-2} (q-q^3)x_1^{(0)} +  \cdots,  \\[2mm]
X_1^2(\z)(x_1^{(1)}) = -\z q x_1^{(2)}
  - \z^3 q^4 x_2^{(2)} + \cdots,   
&X_1^2(\z)(x_2^{(1)}) = \z^{-1} (1-q^2) x_1^{(2)} - \z
(q-q^3)
 x_2^{(2)} + \cdots,  \\[2mm]
X_2^1(\z)(x_1^{(2)}) = \z^{-1} x_1^{(1)} -q \z x_2^{(1)} +
  + \cdots,    
&X_2^1(\z)(x_2^{(2)}) = \z^{-3}(q-q^3)x_1^{(1)} + \z^{-1}
 (1-q^2) x_2^{(1)} + 
\cdots,  \\[2mm]
X_2^3(\z)(x_1^{(2)})= \z^2 q^2 x_1^{(3)} + \cdots,  
&X_2^3(\z)(x_2^{(2)})= -(q-2q^3) x_1^{(3)} + \cdots,  
\\[2mm]
X_3^2(\z)(x_1^{(3)})=  \z^{-2} x_1^{(2)} -q x_2^{(2)} + \cdots, &
\\[-4mm]\er\lb{X1} \end{equation}
and
\vspace*{3mm}
\bev{ll}
\vm X_0^1(\z) (y_1^{(0)})=     \frac{1}{\z} y_1^{(1)} - \z q y_2^{(1)}+\cdots,
& X_0^1(\z) (y_2^{(0)})=   \frac{1}{\z^3}(q-q^3) y_1^{(1)}+ \frac{1}{\z}(1-q^2) y_2^{(1)}+\cdots,\\[2mm]
\vm X_1^0(\z) (y_1^{(1)})=     \z(-q+q^3) y_1^{(0)}+\cdots, \quad
&X_1^0(\z) (y_2^{(1)})=     \frac{1}{\z}(1 -q^2)y_1^{(0)}+\z(-q+q^3) y_2^{(0)}+\cdots,\\[2mm] 
\vm X_1^2(\z) (y_1^{(1)})=     y_1^{(2)}+  \z^2 q^3 y_2^{(2)}+\cdots,\quad 
&X_1^2(\z) (y_2^{(1)})= \frac{1}{\z^2}(q-q^3) y_1^{(2)}-q^2 y_2^{(2)}+\cdots,\\[2mm] 
\vm X_2^1(\z) (y_1^{(2)})=     y_1^{(1)} +\z^2 q^3 y_2^{(1)}+\cdots,\quad
&X_2^1(\z) (y_2^{(2)})= \frac{1}{\z^2}(q-q^3) y_1^{(1)} -q^2 y_2^{(1)}+\cdots,\\[2mm]
\vm X_2^3(\z) (y_1^{(2)})=     \z(-q+q^3)  y_1^{(3)}+\cdots,\quad
&X_2^3(\z) (y_2^{(2)})=     \frac{1}{\z}(1 -q^2) y_1^{(3)}+\z(-q+q^3) y_2^{(3)}+\cdots,\\[2mm]
\vm X_3^2(\z) (y_1^{(3)})=     \frac{1}{\z} y_1^{(2)}- \z q y_2^{(2)}+\cdots,
& X_3^2(\z) (y_2^{(3)})=    \frac{1}{\z^3}(q-q^3) y_1^{(2)}+
\frac{1}{\z}(1-q^2)
y_2^{(2)}+\cdots,
\er\label{Y1}\end{equation}
where each of the coefficients is given to order $q^3$.

Let
us go through the example of how to compute $X_0^1(\z)(x_1^{(0)})$
\big(or rather $X_0^1(\z)(\ga_1^{(0)})$, 
where $x_i^{(a)}=\ga_i^{(a)}(v_{\gl_a^{(3)}})$ -- the $x$'s and $y$'s
appearing in \mref{X1} and \mref{Y1} refer in this context to the associated 
homomorphisms\big).
First of all, it follows from \mref{intexp1} that we have
\bec
(1\ot \Phi_{\gL_0}^{\gL_1}(\z))x_1^{(0)}&=& v_{2\gL_0}\ot \big( 
v_{\gL_1}\ot u_1
- q  f_1 v_{\gL_1}\ot u_0 \z
+q^3 \frac{1}{[2]} f_0 f_1 v_{\gL_1}\ot u_1\z^2
\\&&-q^4  \frac{1}{[2]} f_1 f_0 f_1 v_{\gL_1}\ot
u_0\z^3 +\cdots \big).
\er\label{equal1}\end{equation}
Then, we use the perturbative expression for 
$\Phi_{3\gL_0}^{2\gL_0+\gL_1 V^{(1)} }
(\z) v_{3\gL_0}$,  given in equation  \mref{intexp2} of Appendix \ref{voapp},
from which it follows that
\bec &&(\ga^{(1)}_i\ot 1) \Phi_{3\gL_0}^{2\gL_0+\gL_1 V^{(1)} } (\z)
v_{3\gL_0} = \\[2mm]
&&\quad\quad\quad\quad\quad x_i^{(1)}\ot u_1 
-q f_1 x_i^{(1)}\ot u_0 \z 
+q^5 \frac{1}{[4]+[6]} ([3] f_0 f_1 - f_1 f_0)x_i^{(1)} \ot u_1 +\cdots.\label{equal2}\er\end{equation}
Finally, we compute the coefficients $c_i(\z)$ in the expansion
$X_0^1(\z)(\ga_1^{(0)})=\sli_i c_i(\z) \ga_i^{(1)}$, by
substituting
the right-hand sides of \mref{equal1} and \mref{equal2} into the
defining
equation
\be(1\ot \Phi_{\gL_0}^{\gL_1}(\z))x_1^{(0)} = \sli_i
c_i(\z)(\ga^{(1)}_i\ot 1) \Phi_{3\gL_0}^{2\gL_0+\gL_1 V^{(1)} } (\z)
v_{3\gL_0}.\ee
We find $c_1(\z)=1$ and $c_2(\z)=\z^2q^3$ to order $q^3$.
These are the coefficients given in the first line of \mref{X1}.

In a similar way, we can compute the action of $Z_{00;2}^{12}(\z)$,
which is defined by \mref{cdef2}, i.e., through the commutative
diagram,

\vspace*{10mm}
\begin{center}
\begin{texdraw}
\fontsize{10}{13}\selectfont
\setunitscale 1
\drawdim mm
\arrowheadsize l:2.4 w:1.1 \arrowheadtype t:F
\textref h:C v:C
\move(0 0)
\bsegment
\htext(-1 0){$V(3\gL_0)$}
\esegment
\move(0 20)
\bsegment
\htext(-0.5 0){$V(2\gL_0)\ot V(\gL_0)$}
\esegment
\move(60 20)
\bsegment
\htext(-0.5 0){$V(\gL_1+\gL_0)\ot V^{(1)}_{\z}\ot V(\gL_0)$}
\esegment
\move(120 20)
\bsegment
\htext(-0.5 0){$V(\gL_1+\gL_0)\ot V(\gL_1)\ot V^{(2)}_{\z} $}
\esegment
\move(120 0)
\bsegment
\htext(-0.5 0){$V(2\gL_1+\gL_0)\ot V^{(2)}_\z$}
\esegment
\move(120 5)\avec(120 16)
\move(0 5)\avec(0 16)
\move(10 0)\avec(100 0)
\move(15 20)\avec(35 20)
\move(85 20)\avec(95 20)
\htext(25 25){$\Phi_{2\gL_0}^{\gL_1+\gL_0 V^{(1)}}(\z)$}
\htext(90 25){$\Phi_{\gL_0}^{(1,2)}(\z)$}
\htext(50 5){$\Phi_{3\gL_0}^{2\gL_1+\gL_0 V^{(2)}}(\z)$}
\htext(-5 10){$\ga$}
\htext(130 10){$Z_{00;2}^{12}(\z)(\ga)$}
\end{texdraw}
\end{center}
\nin Making use of equation \mref{intexp12}--\mref{intexp7}, we find 
\be
 Z_{00;2}^{12}(\z) \big(x_1^{(0)}\big)= y_1^{(2)} + \z^2 \frac{[2]}{[4][3]-[2]}y_2^{(2)}+\cdots.\ee

It remains to compute $\iota(x_1^{(0)})$, 
$\iota(X_{0}^{1}(\z)(x_1^{(0)}))$ and 
$\iota(Z_{00;2}^{12}(\z)(x_1^{(0)}))$. Let us go through 
the example of $\iota(x_1^{(0)})$. We must calculate the 
the path coefficients $c(p,x_1^{(0)})$ defined  in
\mref{conv} and \mref{embed3}. As an example, let us do this for the path
$\ket{3}\in P_{2\gL_0,\gL_0;3\gL_0}$. 
First, using \mref{X1}, 
we  calculate the denominator $c^{\ell}(\ket{\rien},x_1^{(0)})$ of
\mref{conv} for several values of $\ell$. In fact
$c^{\ell}(\ket{\rien},x_1^{(0)})=1+O(q^4)$ for all $\ell$, and so since
we are
computing only up to order $q^3$, it never enters the ratio \mref{conv}.
We find the numerator $c^{\ell}(\ket{3},x_1^{(0)})$ has the
following values
\be
c^4(\ket{3},x_1^{(0)})&=& 
\bra{x_1^0} X_1^0(1) X_2^1(1) X_1^2(1)X_0^1(1)\ket{x_1^{(0)}}=-q+2q^3+O(q^5),\nn\\
c^5(\ket{3},x_1^{(0)})&=&
\bra{x_1^{(1)}} X_0^1(1) X_1^0(1) X_2^1(1) X_1^2(1)X_0^1(1)\ket{x_1^{(0)}}=-q+2q^3+O(q^5),\nn\\
\vdots\nn\\
c^{\ell}(\ket{3},x_1^{(0)})&=&-q+2q^3+O(q^5). 
\ee
And hence from \mref{conv}, we have
$c(\ket{3},x_1^{(0)})=-q+2q^3+O(q^4)$.
$c(p,x_1^{(0)})$ of any path  $\ket{p}\in 
\cP_{2\gL_0,\gL_0;3\gL_0}$ can be calculated in a similar way. 
We computed the coefficients of a range of example paths in
$\iota(x_1^{(0)})$ to order $q^3$ (to be precise we considered the
paths $\ket{\rien}$, $\ket{3}$,$\ket{5}$,$\ket{7}$, $\ket{7,5}$, $\ket{9,3}$, $\ket{9,5}$,
$\ket{11,5}$, $\ket{7,5,3}$, $\ket{9,5,3}$, $\ket{11,5,3}$, and
$\ket{5,4,3}$).
We found that the coefficients of each of these paths were equal to those in
expression \mref{vac} for $\vac$.
So our perturbative results are consistent with the identification
$\iota(x_1^{(0)})=\vac$.

In a similar way we have computed the coefficients of certain paths
in $\cP_{0,1;1}$ 
contributing to $\iota(X_{0}^{1}(\z)(x_1^{(0)}))$. 
The notation for paths in $\cP_{0,1;1}$  is such that
$\ket{\rien}=(\cdots\ws  0\ws 1\ws 0\ws 1 \ws 0 \ws 1)$, and
$\ket{2\ell}$ differs from $\ket{\rien}$ only in that $p(2\ell)=2$.
Listing the
path in $\cP_{0,1;1}$
 and then the coefficient $c(p,X_{0}^{1}(\z)(x_1^{(0)}))$, we have to order $q^3$:
\bec 
&\ket{\rien}\quad\quad\quad\quad &\ws\ws 1,\\
&\ket{2}    \quad\quad\quad\quad &-q+(1+\z^2)q^3,\\
&\ket{2\ell}_{\ell>1}\quad\quad\quad\quad &-q+2q^3.
\er\lb{1ket}\end{equation}

Finally, we have computed to order $q^3$ the coefficients for 
certain paths in $\cP_{11;2}$ contributing to
$\iota(Z_{00;2}^{12}(\z)\ket{x_1^{(0)}})$. Here the path notation is
$\ket{\rien}=(\cdots\ws 1\ws 2\ws 1\ws 2 \ws 1 \ws 2)$,
$\ket{2\ell+1}$ differs from it only in that $p(2\ell+1)=0$, and
$\ket{2\ell}$ differs from it only in that $p(2\ell)=3$.
Listing the path and then the coefficient
$c(p,Z_{00;2}^{12}(\z)(x_1^{(0)}))$, we have
\bec 
&\ket{\rien}\quad\quad\quad\quad &\ws\ws 1,\\
&\ket{2}\quad\quad\quad\quad &-q+2q^3,\\
&\ket{2\ell+1}_{\ell>0}\quad\quad\quad\quad &-q+3q^3,\\
&\ket{2\ell}_{\ell>1}\quad\quad\quad\quad
&-q+3q^3.\er\lb{2ket}\end{equation}
\noindent{\bf Step 3}

\nin In this step, we carry out a lattice perturbation theory
calculation of ${_N}X_{0}^{1}(\z)\circ \rho_N \vac$ and
${_N}Z_{00;2}^{12}(\z)\circ \rho_N \circ \vac$. We compare
with the results of Step 2 and hence check the
conjecture \mref{conj}.

First we shall calculate the action of $_N X_0^1(\z)$ on   $\rho_N
\vac$, where $_N X_0^1(\z)$ is defined to be the lattice operator
\mref{latop} in the case when $(m,n)=(1,1)$. Define 
$\ga$, $\gb^{a\pm}$ and $\g^{a\pm}$ to be a factor of 
$\eta(\z^2)/\big(k^{(1,1)}(\z) \eta(\z^{-2})\big)$ times $\bA_3(\z)$, 
$\bB_3^{a\pm}(\z)$ and $\bC_3^{a\pm}(\z)$ respectively. Then as
a series in $q$, we have $\ga(\z)=O(1)$, $\gb^{a\pm}(\z)=O(q)$,
$\g^{a\pm}(\z)=O(1)$.

Let us compute the coefficients of $\ket{\rien}_N$ and
 $\ket{2\ell}_N$
in ${_N}X_0^1(\z) \circ \rho_N \vac$ (where
$\ket{p}_N=\rho_N \ket{p}$, and $\ket{\rien}$ and $\ket{2\ell}$
are as defined above \mref{1ket}). 
We introduce the notation
$\g=(\g^{0+} \g^{1-})^{1/2}$ and define $f_N^{(1,1)}(\z,q)$ by
\begin{equation} f_N^{(1,1)}(\z,q)=\left\{
\br{ll} 
 1+(-1+\z^{-2})q^2+O(q^4) & \hbox{ for }\, N \,\,\hbox{even};\\
 1+O(q^4)&  \hbox{ for }\, N \,\, \hbox{odd}.\er\right.\end{equation}
Then, the coefficients of $\ket{\rien}_N$, $\ket{2}_N$ 
and $\ket{2\ell}_N$ $(\ell>1)$ in ${_N}X_0^1(\z) \circ \rho_N \vac$
when $N$ is large and 
even are given to
order $q^3$ by
\be
&& \g^{N}  -q \ga \gb^{1+} \g^{N-2}(N-2)/2=f_N^{(1,1)}(\z,q),\nn \\[4mm]
&&\ga \gb^{1-} \g^{N-2}+ (-q+2 q^3) \ga^2 \g^{1+} \g^{1-} \g^{N-4}
-q \ga^2 \gb^{1+} \gb^{(1-)} \g^{N-4} (N-4)/2 \nn \\
&& +q^2 \ga^3 \g^{1+} \g^{1-} \gb^{1+} \g^{N-6} (N-2)/2
=f_N^{(1,1)}(\z,q)(-q+(1+\z^2)q^3),\quad\hbox{and}\nn \\[4mm]
&&\ga \gb^{1-} \g^{N-2}+ (-q+2 q^3) \ga^2 \g^{1+} \g^{1-}  \g^{N-4}
+  (-q+2 q^3)\gb^{1-} \gb^{1+} \g^{2-} \g^{0+} \g^{N-4}\nn\\
&&+(-q+2 q^3) \ga^2 \gb^{1+} \gb^{1-} \g^{N-4}(N-6)/2
+ q^2\ga^3 \gb^{1+} \g^{1+} \g^{1-} \g^{N-6} (N-4)/2 \nn\\
&&+ 2q^2 \ga \gb^{1+} \g^{2-} \g^{1+} \g^{N-4} = f_N^{(1,1)}(\z,q) (-q+2q^3)
 \nn\ee
respectively.
When $N$ is large and odd, the three coefficients are
\be
&& \g^{0+} \g^{N-1}-q \ga \gb^{1+} \g^{0+} \g^{N-3}
(N-1)/2=f_N^{(1,1)}(\z,q),
\nn\\[4mm]
&& \ga \gb^{1-} \g^{0+} \g^{N-3}+ (-q+2 q^3) \ga^2 \g^{1+} \g^{N-3}
-q \ga^2 \gb^{1+} \gb^{1-} \g^{0+} \g^{N-5} (N-3)/2 \nn\\
&& + q^2 \ga^3 \g^{1+} \gb^{1+} \g^{N-5} (N-1)/2 
=f_N^{(1,1)}(\z,q)(-q+(1+\z^2)q^3),\quad\hbox{and}\nn\\[4mm]
&& \ga \gb^{1-} \g^{0+} \g^{N-3}
 + (-q+2 q^3)\ga^2 \g^{1+} \g^{N-3}
 + (-q+2 q^3)\gb^{1-} \gb^{1+} \g^{2-} (\g^{0+})^2 \g^{N-5}\nn\\
&&+(-q+2 q^3)\ga^2 \gb^{1+} \gb^{1-} \g^{0+} \g^{N-5} (N-5)/2
+q^2 \ga^3 \gb^{1+} \g^{1+} \g^{N-5} (N-3)/2\nn\\
&&+ 2 q^2 \ga \gb^{1+} \g^{2-} \g^{1+} \g^{0+} \g^{N-5}
=f_N^{(1,1)}(\z,q)(-q+2q^3).\nn \ee
Comparing these coefficients with those of \mref{1ket},
we see that our perturbation theory calculation is consistent with 
conjecture $\mref{conj}$ in the case $(m,n)=(1,1)$.

In order to consider $Z_{00;2}^{12}(\z)\circ \rho_N \circ \vac$
we must first introduce some notation for Boltzmann weights.
There are six independent Boltzmann weights, the formulae for which are given by
\mref{conn1}--\mref{conn2}.

\nin We denote them by
\ben A&=& 
    \wet{0}{2}{1}{3}{\z}{(2,1)}{3}=\wet{3}{1}{2}{0}{\z}{(2,1)}{3},\nn\\
    B_{12}^e&=&
    \wet{1}{1}{2}{2}{\z}{(2,1)}{3}=\wet{2}{2}{1}{1}{\z}{(2,1)}{3},\nn\\
    B_{12}^d&=& 
    \wet{1}{3}{0}{2}{\z}{(2,1)}{3}=\wet{2}{0}{3}{1}{\z}{(2,1)}{3},\nn\\
    C_{10}^e&=&
    \wet{1}{1}{2}{0}{\z}{(2,1)}{3}=\wet{2}{2}{1}{3}{\z}{(2,1)}{3},\nn\\
    C_{12}^e&=&
    \wet{1}{1}{0}{2}{\z}{(2,1)}{3}=\wet{2}{2}{3}{1}{\z}{(2,1)}{3},\nn\\
    C_{01}^d&=&
    \wet{0}{2}{1}{1}{\z}{(2,1)}{3}=\wet{3}{1}{2}{2}{\z}{(2,1)}{3},\nn\\
    C_{12}^d&=&
    \wet{1}{3}{2}{2}{\z}{(2,1)}{3}=\wet{2}{0}{1}{1}{\z}{(2,1)}{3}.\nn
\een
The notation is such that an $e$ superscript implies that the $NW$ and $NE$ 
entries are equal, and a $d$ superscript implies that they are different. The
subscripts give the values of the $(NW,SE)$ pair of entries (for one
of the members of a pair of equal Boltzmann weights). $B$ weights have 0 or 2 horizontal
pairs in which the entries are equal, $C$ weights have one such pair. 
As $q$-series, the $A$ and $C$
weights are $O(1)$ and the $B$ weights are $O(q)$.
Now we compute the $\ket{\rien}_N$, $\ket{2\ell}_N$ and
$\ket{2\ell+1}_N$ 
contributions to $Z_{00;2}^{12}(\z)\circ \rho_N \circ
\vac$ (where $\ket{\rien}$, $\ket{2\ell}$ and  $\ket{2\ell+1}$
are as defined above \mref{2ket}). Let us define
$f^{(2,1)}_{N}(\z,q)$ by
\begin{equation}
f_N^{(2,1)}(\z,q)=\left\{
\br{ll} 
 1+q^3/\z^2+O(q^4) &\hbox{ for }\, N \,\,\hbox{even};\\
 1+q^2/2+O(q^4)&\hbox{ for }\, N \,\,\hbox{odd}.\er\right.\end{equation}
Then, the respective 
coefficients of $\ket{\rien}_N$, $\ket{2}_N$, $\ket{2\ell+1}_N$
$(\ell>0)$ and $\ket{2\ell}_{N}$ $(\ell>1)$ in  $Z_{00;2}^{12}(\z)\circ \rho_N \circ
\vac$ when $N$ is large and even are given  up to order $q^3$ by
\be
&&C^N+ (B_{12}^e)^2 C^{N-2} (-q) (N-2)/2 = f_N^{(2,1)}(\z,q)\nn,\\[2mm]
&&A B^d_{12} C^{N-2}
+A B^e_{12} C^e_{12} C^d_{12} C^{N-4} (-q)= f_N^{(2,1)}(\z,q)(-q+2q^3) \nn,\\[2mm]  
&&C^d_{01} C^e_{12} C^{N-2} (-q+2 q^3) = f_N^{(2,1)}(\z,q)(-q+3q^3)   \nn,\\[2mm]          
&&A B^d_{12} C^{N-2}
+B^d_{12} B^e_{12} C^d_{01} C^e_{10} C^{N-4} (-q)
+A B^e_{12} C^d_{12} C^e_{12} C^{N-4} (-q)  = f_N^{(2,1)}(\z,q)(-q+3q^3).\nn
\ee
When $N$ is large and odd, they are
\be
&&C^e_{10} C^{N-1}+ B^e_{12} B^e_{12} C^e_{10} C^{N-3} (-q) (N-1)/2  =
f_N^{(2,1)}(\z,q) \nn,\\[2mm]       
&&A B^d_{12} C^e_{10} C^{N-3}
+A B^e_{12} C^e_{12} C^{N-3} (-q) = f_N^{(2,1)}(\z,q)(-q+2q^3) \nn,\\[2mm] 
&&C^d_{01} C^e_{12} C^e_{10} C^{N-3} (-q+2 q^3) = f_N^{(2,1)}(\z,q)(-q+3q^3) \nn,\\[2mm] 
&&A B^d_{12} C^e_{10} C^{N-3}                   
+B^d_{12} B^e_{12} C^d_{01} (C^e_{10})^2 C^{N-5} (-q)
+A B^e_{12} C^e_{12} C^{N-3} (-q)  = f_N^{(2,1)}(\z,q)(-q+3q^3).
\nn\ee 
Comparing these coefficients with those of \mref{2ket},
we see that our perturbation theory calculation is consistent with 
the conjecture \mref{conj} in the case $(m,n)=(2,1)$.

\setcounter{equation}{0}
\section{Discussion}
We have constructed a realisation of impurity operators within
the algebraic analysis picture of RSOS models. It
is now a straightforward step to extend the approach 
described in \cite{Fodal93} in order to write down trace expressions
for correlation functions of impurity insertions in these models.
It should also be feasible to construct a free-field realisation of our
impurity operators within the scheme of \cite{LP96}, and to compute
integral formulae for the correlation functions.

Suppose $n=1$. Then if $q$ were equal to $1$, our definition
\mref{cdef1}--\mref{cdef2} of $X$ and $Z_m$ would coincide with 
the coset construction of the Virasoro $q$-primary fields 
$\Phi_{(1,2)}$ and $\Phi_{(m,m+1)}$ respectively. A $q$-Virasoro algebra
was constructed in terms of a free-field realisation in \cite{SKAO},
and in terms of a $q$-coset realisation in \cite{JS97}. A definition
of $q$-primary fields, or $q$-vertex operators, was given 
in \cite{AKMOS} (see also \cite{Kad96}). 
We anticipate that our $X$ and $Z_m$
give a coset construction of the $q$-vertex operators which are 
deformations of $\Phi_{(1,2)}$ and $\Phi_{(m,m+1)}$.

\subsubsection*{Acknowledgements}
I would like to thank Matthias Gaberdiel, Tetsuji Miwa and G\'erard
Watts for their interest and advice. I would also like to acknowledge
funding from EPSRC Advanced Fellowship B/96/AF/2235.
\newpage
\baselineskip=13pt

\baselineskip=17pt
\newpage
\appendix
\setcounter{equation}{0}
\section{Commutation Relations of $\gP^{\gl' V^{(1)}}_{\gl}(\z_1)$ and 
 $\gP^{\gl' V^{(n)}}_{\gl}(\z_2)$ }\lb{weightapp}
In this Appendix, we solve the q-KZ equation in order to
derive the commutation relations for $\gP^{\gl' V^{(1)}}_{\gl}(\z_1)$
and  $\gP^{\gl' V^{(n)}}_{\gl}(\z_2)$. In this way, we arrive at the
explicit expressions for the connection coefficients $C^{(n,1)}_k$ and
$C^{(1,n)}_k$. 

In order to formulate and solve the q-KZ equation it is convenient to
work with a different evaluation module, namely the homogeneous
evaluation module $(V_n)_z$ defined in \cite{idzal93} in terms of
vectors $v^{(n)}_i$, $i\in\{0,1,\cdots,n\}$ (this module is labelled as
$V^{(n)}_z$ in \cite{idzal93}).
The isomorphism between this and the principal evaluation module $V^{(n)}_{\z}$
used elsewhere in this paper is
\bec C_n(\z): & V^{(n)}_{\z} &\curlra (V_n)_z,\\
&u_j^{(n)} &\longmapsto c_j^{(n)} \z^{j} v_j^{(n)},\er\label{evaliso}
\end{equation}
where $c_j^{(n)}={{\qbinom{n}{j}}}^{\!\!\!\!\half} q^{\frac{j}{2}(n-j)}$, and
we identify $\z^2=z$. \,$\qbinom{a}{b}$ is the standard $q$-binomial
coefficient.

We define normalised homogeneous intertwiners 
\bec \tgP_{\gl}^{\mu V_n}(z):&V(\gl)&\lra V(\mu)\ot (V_n)_z \\
 &v_{\gl}&\longmapsto v_{\mu}+ v^{(n)}_j + \cdots , \ws \hbox{where} \ws
\gl=\mu+(n-2j)\bar{\rho},\lb{vonormh}
\er\end{equation} exactly as 
in Section 3.3 of \cite{idzal93}. The relation to the principal
intertwiners defined in Section \ref{intsec} above is
\ben
\gP^{\mu V^{(n)}}_{\gl}(\z) = c^{(n)}_j \z^j C_n(\z)^{-1}
  \tgP_{\gl}^{\mu V_n}(z=\z^2),\ws \hb{where}\ws \gl=\mu+(n-2j)\bar{\rho}.
\een

Now define the matrix element
\ben \Psi^{(m,n)}(z_1/z_2)=\bra{\nu}\tgP^{\nu V_m}_{\mu}(z_2)
\tgP^{\mu V_n}_{\gl}(z_1)
\ket{\gl}\in (V_m)_{z_2}\ot (V_n)_{z_1}.\een
The q-KZ equation for $\Psi^{(m,n)}(z_1/z_2)$ is given by equations
(A.18) and (A.19) of \cite{idzal93}. Let
$\gl=\gl^{(k)}_a$, and define the function $\g(z)$ by
\ben \g(z)&=&\frac{ ( pz q^{1-n};p,q^4)_{\infty} ( pz q^{3+n};p,q^4)_{\infty}}
                    { ( pz q^{5+n};p,q^4)_{\infty}( pz q^{-1-n};p,q^4)_{\infty} }.\een
Then solving 
the q-KZ equation, we
find:\\
when $\mu=\gl_+$,
\bev{l}  \Psi^{(n,1)}(z) = \g(z)
\bigg(
\hhg{-2s(1+j)}{2s(a+1-n+j)}{2s(a+1)}{zpq^{1+n}}  v^{(n)}_j\ot v_1^{(1)} \\
\quad\quad\quad\quad\quad +q^{2(a+1)+n-j}\frac{1-q^{2(j-n)}}{1-q^{2(a+1)}}
\hhg{1-2s(1+j)}{2s(a+1-n+j)}{1+2s(a+1)}{zpq^{1+n}} 
v^{(n)}_{j+1} \ot v_0^{(1)}
\bigg).
\er\lb{matrix1}\end{equation}

\noindent When $\mu=\gl_-$,
\bev{l}  \Psi^{(n,1)}(z) = \g(z)
\bigg(
\hhg{2s(-n+j-1)}{1-2s(a+j+1)}{1-2s(a+1)}{zpq^{1+n}}  v^{(n)}_j\ot v_0^{(1)}\\
\quad\quad\quad\quad\quad +z p q^{-2(a+1)+j}\frac{1-q^{-2j}}{1-p q^{-2(a+1)}}
\hhg{1+2s(-n+j-1)}{1-2s(a+j+1)}{2-2s(a+1)}{zpq^{1+n}}
v^{(n)}_{j-1}\ot v_1^{(1)}
\bigg).
\er\lb{matrix2}\end{equation}

\nin When $\nu=\mu_+$,
\bev{l}  \Psi^{(1,n)}(z) = \g(z)
\bigg(
\hhg{-2s(1+j)}{1-2s(a+j+2)}{1-2s(a-n+2j+2)}{zpq^{1+n}}v_1^{(1)}\ot v^{(n)}_j\\
\quad\quad\quad\quad\quad +z q^{j-n}\frac{1-q^{2(n-j)}}{1-p^{-1}q^{2(a+2-n+2j)}}
\hhg{1-2s(1+j)}{1-2s(a+j+2)}{2-2s(a-n+2j+2)}{zpq^{1+n}} 
 v_0^{(1)}\ot v^{(n)}_{j+1}
\bigg).
\er\lb{matrix3}\end{equation}

\nin When $\nu=\mu_-$,
\bev{l} \Psi^{(1,n)}(z) = \g(z)
\bigg(
\hhg{2s(-n+j-1)}{2s(a-n+j)}{2s(a-n+2j)}{z pq^{1+n}}v_0^{(1)}\ot  v^{(n)}_j\\
\quad\quad\quad\quad\quad +q^{-j}\frac{1-q^{2j}}{1-q^{-2(a-n+2j)}}
\hhg{1+2s(-n+j-1)}{2s(a-n+j)}{1+2s(a-n+2j)}{zpq^{1+n}} 
 v_1^{(1)}\ot v^{(n)}_{j-1}
\bigg).
\er\label{matrix4}\end{equation}

In all cases, $j$ is determined uniquely by the requirement that 
weight($\tgPs(z)$)=$\gl-\nu$. The function $\phi$ is the basic
hypergeometric series 
\ben \hhg{\ga}{\gb}{\g}{z}=\fullhhg{p^\ga}{p^\gb}{p^\g}{p}{z}=
\sli_{n=0}^{\infty}\frac{(p^\ga;p)_n (p^\gb;p)_n}{(p^\g;p)_n (p;p)_n }z^n.\een
The normalisation of the first term in each of \mref{matrix1}--\mref{matrix4} 
is fixed by \mref{vonormh}. The normalisation of the second
term follows from the $q-$KZ equation, and is computed by making use of
the identities
{\small
\ben (1-zp^\ga)\hhg{\ga}{\gb}{\g}{pz}-(1-z)\hhg{\ga}{\gb}{\g}{z} &=&    
    z(p^\gb-p^\g)\frac{(1-p^\ga)}{(1-p^\g)} \hhg{1+\ga}{\gb}{1+\g}{pz},\nn\\[2mm]
    (1-zp^{\ga+\gb+\g})\hhg{\ga}{\gb}{\g}{pz}-(1-zp^{\gb-\g})\hhg{\ga}{\gb}{\g}{z} &=& 
    -z(1-p^{\gb-\g})\frac{(1-p^\ga)}{(1-p^\g)} \hhg{1+\ga}{\gb}{1+\g}{z}.\nn\een
}

Given \mref{matrix1}-\mref{matrix4}, the explicit form of the homogeneous
$R$-matrix $\bar{R}^{(1,n)}(z)$ given in Section 3.2 of \cite{idzal93}, 
the connection formula (B.8) of \cite{idzal93}, and the isomorphism
\mref{evaliso}, one can then compute the connection coefficients
$C^{(n,1)}_k$ and $C^{(1,n)}_k$ defined in \mref{comm1}. We find 

 \be
\cet{\gl}{\mu}{\mu'}{\nu}{\z}{(n,1)}{k}=
\frac{1}{\gk^{(n,1)}(\z)} \bcet{\gl}{\mu}{\mu'}{\nu}{\z}{(n,1)}{k},
\lb{cn1}\ee
where
\begin{equation}\br{l} \bcet{\gl}{\nu_+}{\gl_+}{\nu}{\z}{(n,1)}{k}= 
\z q^{\half (n-2j+1)}\frac{[n-j+1]^{\half}}{[j]^{\half}}  
\frac{\eta(\z^2)}{\eta(\z^{-2})}
\frac{\gG_p(2s(a+2j-n))\gG_p(1-2s(a+1))\gT_p(pq^{-2(a+j)+n-1}\z^2)  }
     {\gG_p(1+2s(j-1-n))\gG_p(2sj)\gT_p(q^{1+n}\z^2)  },
\\[2mm]
    \bcet{\gl}{\nu_+}{\gl_-}{\nu}{\z}{(n,1)}{k}=  q^{j}\frac{\eta(\z^2)}{\eta(\z^{-2})} 
\frac{\gG_p(2s(a+2j-n))\gG_p(2s(a+1))\gT_p(q^{-2j+n+1}\z^2)  }
     {\gG_p(2s(a+j-n))\gG_p(2s(a+j+1)\gT_p(q^{1+n}\z^2)  },
\er \lb{conn1}\end{equation}
\nin with $j$ given by $\nu_+ +(n-2j)\bar{\rho}=\gl$,
and 
\begin{equation}\br{l}
 \bcet{\gl}{\nu_-}{\gl_+}{\nu}{\z}{(n,1)}{k}=  q^{n-j}\frac{\eta(\z^2)}{\eta(\z^{-2})}
\frac{\gG_p(1-2s(a+2j-n+2))\gG_p(1-2s(a+1))\gT_p(q^{2j+1-n}\z^2)  }
     {\gG_p(1-2s(a+j+2))\gG_p(1-2s(a+j+1-n)\gT_p(q^{1+n}\z^2)  },
\\[2mm]
    \bcet{\gl}{\nu_-}{\gl_-}{\nu}{\z}{(n,1)}{k}= 
\z q^{\half (2j-n+1)} \frac{[j+1]^{\half}}{[n-j]^{\half}}
\frac{\eta(\z^2)}{\eta(\z^{-2})}
\frac{\gG_p(1-2s(a+2j-n+2))\gG_p(2s(a+1))\gT_p(q^{2a+2j+3-n}\z^2)  }
     {\gG_p(1-2s(1+j))\gG_p(2s(n-j)\gT_p(q^{1+n}\z^2)  },
\er\lb{conn2}\end{equation}
with $j$ given by $\nu_- +(n-2j)\bar{\rho}=\gl$.
The functions $\gG_p$ are $\gT_p$ are defined as usual by
\be \gG_p(z)=\frac{(p;p)_{\infty}}{(p^z;p)_{\infty}}(1-p)^{1-z},\quad 
\gT_p(z)=(p;p)_{\infty} (z;p)_{\infty} (pz^{-1};p)_{\infty},\lb{fndef}\ee
and  $\eta(\z)$ is defined by
\be \eta(z)= \frac{(pzq^{1+n};p,q^4)_{\infty}
  (pzq^{3-n};p,q^4)_{\infty}}
{(pzq^{1-n};p,q^4)_{\infty} (pzq^{3+n};p,q^4)_{\infty}}, \ws{\rm
  with} \ws (a;b,c)_{\infty}\equiv \pl_{n_1,n_2=0}^{\infty}(1-a\, b^{n_1} c^{n_2}).
\label{eta}\ee

We also find 
\be 
\cet{\gl}{\mu}{\mu'}{\nu}{\z}{(1,n)}{k}=\cet{\nu}{\mu}{\mu'}{\gl}{\z}{(n,1)}{k},\lb{cnf}\ee
such that the Boltzmann weights of Section \mref{lattsec} are given by
\ben \wet{\gl}{\mu}{\mu'}{\nu}{\z}{(n,1)}{k}=\cet{\gl}{\mu}{\mu'}{\nu}{\z}{(n,1)}{k}.\een
\newpage
\setcounter{equation}{0}
\section{Commutation Relations of $\gP^{(n,n+k)}_{\gl}(\z)$}
In this Appendix we give a proof of the commutation relations
\be
 R^{(n+k,n+k)}(\z)
  \gP^{(n,n+k)}_{\gs(\gl)}(\z_1)\gP^{(n,n+k)}_{\gl}(\z_2)= 
  \gP^{(n,n+k)}_{\s(\gl)}(\z_2) \gP^{(n,n+k)}_{\gl}(\z_1)
  R^{(n,n)}(\z),\label{pp}\ee
where $\z=\z_1/\z_2$. 
The proof will be inductive on the level $k$.

\mref{pp}
is shown for $k=1$ in \cite{HKMW98b}, and we make the assumption that it is
true for $k=\ell-1$.
Let $\gl=\mu+\gL_i$ and consider
\be
R^{(n+\ell,n+\ell)}(\z) 
\big(\gP^{(n+\ell-1,n+\ell)}_{\gL_{1-i}}(\z_1)  \gP^{(n,n+\ell-1)}_{\gs(\mu)}(\z_1)\big)
\big(\gP^{(n+\ell-1,n+\ell)}_{\gL_{i}}(\z_2)
\gP^{(n,n+\ell-1)}_{\mu}(\z_2)\big),
\lb{st1}\ee
which is an intertwiner $V^{(n)}_{\z_1}\ot V^{(n)}_{\z_2} \ot
V(\mu)\ot V(\gL_i) \to V(\mu)\ot V(\gL_i) \ot V^{(n+\ell)}_{\z_2}\ot
V^{(n+\ell)}_{\z_1}.$ 
Since $\gP^{(n,n+\ell-1)}_{\gs(\mu)}(\z_1)$ and
$\gP^{(n+\ell-1,n+\ell)}_{\gL_{i}}(\z_2)$  act on
different spaces, they commute. So
\mref{st1} is equal to
\be R^{(n+\ell,n+\ell)}(\z) 
\gP^{(n+\ell-1,n+\ell)}_{\gL_{1-i}}(\z_1) 
\gP^{(n+\ell-1,n+\ell)}_{\gL_{i}}(\z_2)\gP^{(n,n+\ell-1)}_{\gs(\mu)}(\z_1)
\gP^{(n,n+\ell-1)}_{\mu}(\z_2).
\nn\ee
Using \mref{pp} when $k=1$, this is equal to
\be 
\gP^{(n+\ell-1,n+\ell)}_{\gL_{1-i}}(\z_2) 
\gP^{(n+\ell-1,n+\ell)}_{\gL_{i}}(\z_1)
R^{(n+\ell-1,n+\ell-1)}(\z) 
\gP^{(n,n+\ell-1)}_{\gs(\mu)}(\z_1)
\gP^{(n,n+\ell-1)}_{\mu}(\z_2).
\nn\ee
Now using \mref{pp} when $k=\ell-1$, this becomes
\be 
\gP^{(n+\ell-1,n+\ell)}_{\gL_{1-i}}(\z_2) 
\gP^{(n+\ell-1,n+\ell)}_{\gL_{i}}(\z_1)
\gP^{(n,n+\ell-1)}_{\gs(\mu)}(\z_2)
\gP^{(n,n+\ell-1)}_{\mu}(\z_1)
R^{(n,n)}(\z).
\nn\ee
Using the commutativity of $\gP^{(n+\ell-1,n+\ell)}_{\gL_{i}}(\z_1)$ and
$\gP^{(n,n+\ell-1)}_{\gs(\mu)}(\z_2)$ we thus arrive at the equality
\bev{l}
R^{(n+\ell,n+\ell)}(\z) 
\big(\gP^{(n+\ell-1,n+\ell)}_{\gL_{1-i}}(\z_1)  \gP^{(n,n+\ell-1)}_{\gs(\mu)}(\z_1)\big)
\big(\gP^{(n+\ell-1,n+\ell)}_{\gL_{i}}(\z_2)
\gP^{(n,n+\ell-1)}_{\mu}(\z_2)\big)\\
=\big(\gP^{(n+\ell-1,n+\ell)}_{\gL_{1-i}}(\z_2) 
\gP^{(n,n+\ell-1)}_{\gs(\mu)}(\z_2)\big)
\big(\gP^{(n+\ell-1,n+\ell)}_{\gL_{i}}(\z_1)
\gP^{(n,n+\ell-1)}_{\mu}(\z_1)\big)
R^{(n,n)}(\z). 
\lb{fstep}\er\end{equation}

It is shown in \cite{HKMW98b} that $\gP^{(n+\ell-1,n+\ell)}_{\gL_{i}}(\z)
\gP^{(n,n+\ell-1)}_{\mu}(\z)=(\gP^{(n+\ell-1,n+\ell)}_{\gl}(\z)\ot \id)$
when restricted to $V(\gl)\ot\gO_{\mu,\gL_i;\gl}$ with
$\gl=\mu+\gL_i$. Hence restricting \mref{fstep} to
$V(\gl)\ot\gO_{\mu,\gL_i;\gl}$ gives \mref{pp} with $k=\ell$. This
completes the proof.

\newpage
\setcounter{equation}{0}
\section{The Perturbative Action of Intertwiners}\label{voapp}
In this Appendix, we list the perturbative action of the intertwiners
used in Section \ref{pertsec}. We have
\be 
\Phi_{\gL_0}^{\gL_1 V^{(1)}}(\z)v_{\gL_0}&=&
 v_{\gL_1}\ot u^{(1)}_1
- q  f_1 v_{\gL_1}\ot u^{(1)}_0 \z
+\frac{q^3}{[2]} f_0 f_1 v_{\gL_1}\ot u^{(1)}_1\z^2\nn\\
&&-\frac{q^4}{[2]} f_1 f_0 f_1 v_{\gL_1}\ot
u^{(1)}_0\z^3
+\cdots,\lb{intexp1}\\[3mm]
\Phi_{2\gL_0}^{\gL_0+\gL_1V^{(1)}}(\z) v_{2\gL_0}&=& v_{\gL_0+\gL_1}\ot
u^{(1)}_1-qf_1 v_{\gL_0+\gL_1} \ot
u^{(1)}_0\z\nn\\&&+\frac{q^4}{1-[3]^2}(f_1f_0-[3]f_0f_1) v_{\gL_0+\gL_1}\ot
u^{(1)}_1 \z^2 +\cdots,\lb{intexp12}\\[3mm]
\Phi_{3\gL_0}^{2\gL_0+\gL_1 V^{(1)} } (\z) v_{3\gL_0}&=&
v_{2\gL_0+\gL_1}\ot u^{(1)}_1 
-q f_1 v_{\gL_{2\gL_0}+\gL_1}\ot u^{(1)}_0 \z\nn \\
&&+\frac{q^5}{[4]+[6]} ([3] f_0 f_1 - f_1 f_0) v_{\gL_{2\gL_0}+\gL_1}\ot u^{(1)}_1
\z^2 +\cdots,
\lb{intexp2}\\[3mm]
\Phi^{3\gL_0 V^{(1)} }_{2\gL_0+\gL_1 } (\z)v_{2\gL_0+\gL_1}&=&
v_{3\gL_0}\ot u^{(1)}_0 
-\frac{q^3}{[3]} f_0 v_{3\gL_0}\ot u^{(1)}_1 \z 
+\frac{q^5}{[2][3]} f_1 f_0 v_{3 \gL_0}\ot u^{(1)}_0 \z^2 + \cdots,
\lb{intexp3}\\[3mm]
\Phi^{\gL_0+2\gL_1 V^{(1)} }_{2\gL_0+\gL_1 } (\z)v_{2\gL_0+\gL_1}&=&
v_{\gL_0+2\gL_1}\ot u^{(1)}_1 
-\frac{q^2}{[2]} f_1 v_{\gL_0+2\gL_1}\ot u^{(1)}_0 \z \nn\\&& 
+\frac{q^5}{[2]([3][4]-[2])} ([4] f_0 f_1 -[2] f_1 f_0)
v_{\gL_0+2\gL_1}\ot
 u^{(1)}_1 \z^2 +\cdots,
\lb{intexp4}\\
\Phi_{3\gL_0}^{2\gL_1+\gL_0 V^{(2)}}(\z) v_{3\gL_0}&=&
v_{2\gL_1+\gL_0}\ot u_2^{(2)} -\frac{q^{3/2}}{[2]^{1/2}} f_1 v_{2\gL_1+\gL_0}\ot
u_1^{(2)}\z
+\frac{q^2}{[2]}f_1^2 v_{2\gL_1+\gL_0}\ot u_0^{(2)}\z^2 \nn\\&&+ 
\frac{q^5}{[4][3]-[2]}([4]f_0f_1-[2]f_1 f_0)v_{2\gL_1+\gL_0}\ot u_2^{(2)}\z^2+\cdots
,\lb{intexp5}\\
\Phi_{\gL_0}^{(1,2)}(\z)(u^{(1)}_1\ot v_{\gL_0})&=&
v_{\gL_1}\ot u_2^{(2)}-\frac{q^{3/2}}{[2]^{1/2}} f_1 v_{\gL_1}\ot
u_1^{(2)}\z
+\frac{q^4}{[2]} f_0f_1v_{\gL_1}\ot u_{2}^{(2)}\z^2\cdots
,\lb{intexp6}\\
\Phi_{\gL_0}^{(1,2)}(\z)(u^{(1)}_0\ot v_{\gL_0})&=&
\frac{q^{-1/2}}{[2]^{1/2}} v_{\gL_1}\ot u_1^{(2)} -q f_1 v_{\gL_1} \ot
u_{0}^{(2)}\z+\frac{q^{7/2}}{[2]^{3/2}} f_0 f_1 v_{\gL_1} \ot
u_1^{(2)}\z^2 +\cdots.\lb{intexp7}
\ee

All other intertwiners we need are given by a $(f_i,\gL_j,u^{(n)}_{\ell})\leftrightarrow
(f_{1-i},\gL_{1-j},u^{(n)}_{n-\ell})$ symmetry, for example the expansion
\ben 
\Phi_{\gL_1}^{\gL_0 V^{(1)}}(\z)v_{\gL_1}&=&
 v_{\gL_0}\ot u^{(1)}_0
- q  f_0 v_{\gL_0}\ot u^{(1)}_1 \z
+q^3 \frac{1}{[2]} f_1 f_0 v_{\gL_0}\ot u^{(1)}_0\z^2\\
&& -q^4  \frac{1}{[2]} f_0 f_1 f_0 v_{\gL_0}\ot
u^{(1)}_1\z^3
+\cdots\nn,\een follows from \mref{intexp1} under this symmetry. This symmetry
is one of the benefits of using a principal evaluation module.
\end{document}